\documentclass[structabstract,longauth]{aa}  
\usepackage{graphicx}
\usepackage[latin1]{inputenc}
\usepackage{amsmath}
\usepackage{amsfonts}
\usepackage{amssymb}
\usepackage{natbib}
\usepackage{txfonts}
\usepackage{morefloats}
\usepackage{subfig}

\newcommand{\fchartText}[4]{#1: Finding chart constructed from the reference image marking the positions of the variable stars. #2North is up and east is to the right. The cluster image is $\sim$ $#3^{\prime\prime}$ by $#4^{\prime\prime}$.}

\newcommand{\varTableText}[2]{The celestial coordinates correspond to the epoch of the reference image, which is the HJD $\sim$#1 d. Epochs are (HJD-2450000). $< I >$ denotes mean $I$ magnitude and $A_{i'+z'}$ are the amplitudes found in our special $i'+z'$ filter. The blend column describes whether the star is blended with (i) brighter star(s), (ii) star(s) of similar magnitude, or (iii) fainter/no star(s). #2}

\newcommand{\VarFigText}[2]{#1: Light curves for #2 variables in our FoV. Red triangles are 2013 data and blue circles are 2014 data. Error bars are plotted but are smaller than the data symbols in many cases. Light curves with confirmed periods are phased. For those variables without periods the x-axis refers to (HJD - 2450000), and the dashed line indicates that the period from HJD 2456570 to 2456760 has been removed from the plot, as no observations were performed during this time range. }

\graphicspath{{figs/}}

\begin{document}

   \title{Searching for variable stars in the cores of five metal-rich globular clusters using EMCCD observations\thanks{Based on data collected by MiNDSTEp with the Danish 1.54 m telescope}$^{,}$\thanks{The full light curves presented in this paper (Table \ref{table:all_vars}) are only available in electronic form at the CDS via anonymous ftp to cdsarc.u-strasbg.fr (130.79.128.5) or via http://cdsweb.u-strasbg.fr/cgi-bin/qcat?J/A+A/573/A103}}


   \author{Jesper~Skottfelt  \inst{1,2} \thanks{email: skottfelt@nbi.dk}
      \and D.~M.~Bramich \inst{3} 
      \and R.~Figuera~Jaimes \inst{4,5}
      \and U.~G.~J\o{}rgensen \inst{1,2} 
 	  \and N.~Kains \inst{6,4}
 	  \and A.~Arellano~Ferro \inst{7} \and \\
      and \\
	       K.~A.~Alsubai \inst{3}
	  \and V.~Bozza \inst{8,9}
	  \and S.~Calchi~Novati \inst{10,8,11} \thanks{Sagan visiting fellow}
	  \and S.~Ciceri \inst{12}
	  \and G.~D'Ago \inst{8,9}
	  \and M.~Dominik \inst{5} \thanks{Royal Society University Research Fellow}
	  \and P.~Galianni \inst{5}
	  \and S.-H.~Gu  \inst{13,14}
 	  \and K.~B.~W~Harps\o{}e \inst{1,2}
	  \and T.~Haugb\o{}lle \inst{2,1}
	  \and T.~C.~Hinse \inst{15}
	  \and M.~Hundertmark \inst{1,2,5}
	  \and D.~Juncher \inst{1,2}
	  \and H.~Korhonen \inst{16,1,2}
	  \and C.~Liebig \inst{5}
	  \and L.~Mancini \inst{12}
	  \and A.~Popovas \inst{1,2}
	  \and M.~Rabus \inst{17,12}
	  \and S.~Rahvar \inst{18}
	  \and G.~Scarpetta \inst{10,8,9}
	  \and R.~W.~Schmidt \inst{19}
	  \and C.~Snodgrass \inst{20,21}
	  \and J.~Southworth \inst{22}
	  \and D.~Starkey \inst{5}
	  \and R.~A.~Street \inst{23}
	  \and J.~Surdej \inst{24}
	  \and X.-B.~Wang \inst{13,14}
	  \and O.~Wertz \inst{24}\\
	  (The MiNDSTEp Consortium)}

   \institute{Niels Bohr Institute, University of Copenhagen, Juliane Maries Vej 30, 2100 K\o{}benhavn \O{}, Denmark 
    \and Centre for Star and Planet Formation, Natural History Museum, University of Copenhagen, \O{}stervoldgade 5-7, 1350 K\o{}benhavn~K, Denmark 
    \and Qatar Environment and Energy Research Institute, Qatar Foundation, Tornado Tower, Floor 19, P.O. Box 5825, Doha, Qatar 
    \and European Southern Observatory, Karl-Schwarzschild-Stra\ss{}e 2, 85748 Garching bei M\"{u}nchen, Germany 
    \and SUPA, School of Physics and Astronomy, University of St. Andrews, North Haugh, St Andrews, KY16 9SS, United Kingdom.
	\and Space Telescope Science Institute, 3700 San Martin Drive, Baltimore, MD 21218, United States of America 
	\and Instituto de Astronom\'ia Universidad Nacional Aut\'onomade Mexico 
    \and Dipartimento di Fisica ''E. R. Caianiello'', Universit\`a di Salerno, Via Giovanni Paolo II 132, 84084-Fisciano (SA), Italy 
    \and Istituto Nazionale di Fisica Nucleare, Sezione di Napoli, Napoli, Italy 
	\and NASA Exoplanet Science Institute, MS 100-22, California Institute of Technology, Pasadena CA 91125 
    \and Istituto Internazionale per gli Alti Studi Scientifici (IIASS), 84019 Vietri Sul Mare (SA), Italy 
	\and Max Planck Institute for Astronomy, K\"onigstuhl 17, 69117 Heidelberg, Germany 
	\and Yunnan Observatories, Chinese Academy of Sciences, Kunming 650011, China  
	\and Key Laboratory for the Structure and Evolution of Celestial Objects, Chinese Academy of Sciences, Kunming 650011, China 
	\and Korea Astronomy and Space Science Institute, Daejeon 305-348, Republic of Korea 
	\and Finnish Centre for Astronomy with ESO (FINCA), University of Turku, V{\"a}is{\"a}l{\"a}ntie 20, FI-21500 Piikki{\"o}, Finland 
	\and Instituto de Astrof\'isica, Facultad de F\'isica, Pontificia Universidad Cat\'olica de Chile, Av. Vicu\~na Mackenna 4860, 7820436 Macul, Santiago, Chile 
	\and Department of Physics, Sharif University of Technology, P. O. Box 11155-9161 Tehran, Iran 
	\and Astronomisches Rechen-Institut, Zentrum f\"ur Astronomie der Universit\"at Heidelberg, M\"onchhofstr. 12-14, 69120 Heidelberg, Germany 
    \and Planetary and Space Sciences, Department of Physical Sciences, The Open University, Milton Keynes, MK7 6AA, UK 
    \and Max-Planck-Institute for Solar System Research, Justus-von-Liebig-Weg 3, 37077 G\"ottingen, Germany 
	\and Astrophysics Group, Keele University, Staffordshire, ST5 5BG, UK 
	\and Las Cumbres Observatory Global Telescope Network, 6740 Cortona Drive, Suite 102, Goleta, CA 93117, USA 
	\and Institut d'Astrophysique et de G\'eophysique, Universit\'e de Li\`ege, All\'ee du 6 Ao\^ut, B\^at. B5c, 4000 Li\`ege, Belgium 
    }	
	
   \date{\today}

\date{Received 12 September 2014 /
 	  Accepted 31 October 2014}

 
  \abstract
 {}
{In this paper, we present the analysis of time-series observations from 2013 and 2014 of five metal-rich ([Fe/H] $>$ -1) globular clusters: NGC~6388, NGC~6441, NGC~6528, NGC~6638, and NGC~6652. The data have been used to perform a census of the variable stars in the central parts of these clusters. }
{The observations were made with the electron-multiplying CCD (EMCCD) camera at the Danish 1.54m Telescope at La Silla, Chile, and they were analysed using difference image analysis (DIA) to obtain high-precision light curves of the variable stars. } 
{It was possible to identify and classify all of the previously known or suspected variable stars in the central regions of the five clusters. Furthermore, we were able to identify and, in most cases, classify 48, 49, 7, 8, and 2 previously unknown variables in NGC~6388, NGC~6441, NGC~6528, NGC~6638, and NGC~6652, respectively. 
Especially interesting is the case of NGC~6441, for which the variable star population of about 150 stars has been thoroughly examined by previous studies, including a Hubble Space Telescope study. 
In this paper we are able to present 49 new variable stars for this cluster, of which one (possibly two) are RR Lyrae stars, two are W Virginis stars, and the rest are long-period semi-regular or irregular variables on the red giant branch. We have also detected the first double-mode RR Lyrae in the cluster. } 
  {}


   \authorrunning{Skottfelt et al.}
   \titlerunning{Variable stars in metal-rich globular clusters}
   \maketitle


\section{Introduction} \label{sec:intro}
Galactic globular clusters represent some of the oldest stellar populations in the Galaxy and their study can provide important insight into the formation and early evolution of the Galaxy.
The stars in globular clusters are believed to have formed roughly at the same time (although there is also mounting evidence that some clusters have formed via several episodes of star formation), from the same primordial material, and with the same composition. 
This should lead to a homogeneity in certain fundamental properties of the stars within each cluster, but with differences between the clusters.
Knowledge about these properties, such as metallicity, age, and kinematics, are therefore important. 
One way to obtain better constraints on some of these physical parameters is to study the population of variable stars, especially RR Lyrae stars.

In \cite{Skottfelt2013} two new variable stars were discovered in the otherwise well-studied globular cluster NGC~6981. 
The authors made use of electron multiplying CCD (EMCCD) data, which made it possible to retrieve the photometry in an area that, in a conventional CCD, was affected by the saturation of a bright star.

Saturation of bright stars in conventional CCD observations is often hard to avoid when a reasonable signal-to-noise ratio (S/N) for the fainter objects is required. 
When using difference image analysis (DIA), which is currently the best method of extracting precise photometry in crowded regions, saturated pixels affect the nearby pixels during the convolution of the reference image. 
It is thus impossible to perform photometric measurements using DIA near a saturated star in conventional CCD observations, and this affects the completeness of variability studies in crowded regions.

An EMCCD is a conventional CCD with an extended readout register where the signal is amplified by impact ionisation before it is read out. 
The readout noise is thus negligible compared to the signal, even at very high readout speeds (10-100 frames/s), which makes high frame-rate imaging feasible. 
By shifting and adding the individual frames to counteract the blurring effects of atmospheric turbulence, it is possible to obtain very high spatial resolution, something that has been described in numerous articles \citep[e.g.][]{Mackay2004,Law2006A}. 

The use of EMCCD data for precise time-series photometry, however, is a new area of investigation, and the applications are just starting to be explored. 
High frame-rate imaging makes it possible to observe much brighter stars without saturating the CCD, and the required S/N for the object of interest can be achieved by combining the individual exposures into a stacked image at a later stage. 
Combining many exposures thus gives a very wide dynamical range, and the possibility to perform high-precision photometry for both bright and faint stars. 
It should be noted that EMCCD exposures need to be calibrated in a different way than conventional CCD imaging data \citep{Harpsoe2012}.

The DIA technique, first introduced by \citet{Alard1998}, has been improved by revisions to the algorithm presented by \citet{Bramich2008} \citep[see also][]{Bramich2013} and is the optimal way to perform photometry with EMCCD data in crowded fields.
This method uses a flexible discrete-pixel kernel model instead of modelling the kernel as a combination of Gaussian basis functions, and it has been shown to give improved photometric precision even in very crowded regions \citep{Albrow2009}. The method is also especially adept at modelling images with PSFs that are not approximated well by a Gaussian or a Moffat profile.\\

As part of a series of papers on detecting and characterising the variable stars in globular clusters \citep[e.g.][]{Kains2013, Kains2012, FigueraJaimes2013, Skottfelt2013,  ArellanoFerro2013,Bramich2011},
we present our analysis of EMCCD observations of five metal-rich ([Fe/H] $>$ -1) globular clusters. The clusters are listed in Table~\ref{table:NGCs} along with their celestial coordinates, metallicity,
and central concentration parameter.


\begin{table}
 \caption{Coordinates and physical parameters of the five clusters}
 \label{table:NGCs}
 \centering
 \begin{tabular}{ccccccc}
 \hline\hline
 Cluster & RA (J2000.0) & Dec. (J2000.0) & [Fe/H] & $c$ \\
  \hline
NGC~6388  &  17:36:17.23  &  -44:44:07.8 & $-0.55 \pm 0.15$  &  1.75  \\
NGC~6441  &  17:50:13.06  &  -37:03:05.2 & $-0.46 \pm 0.06$  &  1.74  \\ 
NGC~6528  &  18:04:49.64  &  -30:03:22.6 & $-0.12 \pm 0.24$  &  1.5   \\ 
NGC~6638  &  18:30:56.10  &  -25:29:50.9 & $-0.95 \pm 0.13$  &  1.33  \\ 
NGC~6652  &  18:35:45.63  &  -32:59:26.6 & $-0.81 \pm 0.17$  &  1.8   \\ 
 \hline
 \end{tabular}
\tablefoot{
The celestial coordinates are taken from \citet{Harris1996} (2010 edition). Col. 4 gives the metallicity of the cluster \citep[from ][]{Roediger2014}. Col. 5 gives the central concentration $c=\log_{10}(r_t/r_c)$, where $r_t$ and $r_c$ are the tidal and core radii, respectively \citep[from][(2010 edition)]{Harris1996}.
}
\end{table}

The variable star populations of all five clusters have been examined previously, but the dates, methods, and completeness of these studies varies a lot. 

NGC~6388 and NGC~6441 have been thoroughly studied. The observations made by \citet{Pritzl2001, Pritzl2002}, where PSF-fitting photometric methods were used, were re-analysed by \citet{Corwin2006} using image subtraction methods, and this revealed a number of new variables, especially in the crowded central regions. 
NGC~6441 was furthermore included in a Hubble Space Telescope (HST) snapshot program where many new variables were detected \citep{Pritzl2003}.

Although NGC~6528 has been heavily studied due to its very high metal abundance, no variable stars have been found to date.

Since the central region of NGC~6638 is not heavily crowded, \citet{Rutily1977} were able to find several variable stars very close to the centre using only photographic plates. They were, however, not able to determine their periods.

\citet{Hazen1989} found several variables inside the tidal radius of NGC~6652, but none in the relatively crowded centre. 

Due to the relatively small field-of-view (FoV) of the instrument we used, $45\arcsec \times 45\arcsec$, we only observed the central part of these clusters. 
These are the regions of the clusters which benefit the most from the gain in spatial resolution afforded by EMCCD observations.
Therefore, this study cannot be taken as a complete census of the variable star population in each cluster, but only of their crowded central regions.
Another important part of the article is to show that the techniques introduced in \citet{Skottfelt2013}, i.e. using the very high spatial resolution and high-precision photometry of EMCCDs to locate variable stars in crowded fields, can be performed routinely on a much larger scale.

In Sec. \ref{sec:data} we describe our observations and how they were reduced. In Sec. \ref{sec:results} our results for each of the five clusters are presented and Sec. \ref{sec:discussion} contains a discussion of these results. Finally, in Sec. \ref{sec:conclusion} a summary of our results is given.

\section{Data and reductions} \label{sec:data}

\subsection{Observations}
The observations were carried out with the EMCCD instrument at the Danish 1.54m telescope at La Silla, Chile \citep{Skottfelt2015b}. 
The EMCCD instrument consists of a single Andor Technology iXon+ model 897 EMCCD camera which has an imaging area of $512 \times 512$ $16 \mu$m pixels. 
The pixel scale is $\sim 0\farcs09$, and the FoV is therefore $45\arcsec \times 45\arcsec$. 
For these observations a frame-rate of 10 Hz and an EM gain of 1/300 photons$/e^-$ was used. 
The small FoV means that we were limited to targeting the crowded central regions of the clusters. 
The camera sits behind a dichroic mirror that acts as a long-pass filter with a cut-on wavelength of 650~nm. The cut-off wavelength is determined by the sensitivity of the camera which drops to zero at 1050~nm over about 250~nm. The filter thus corresponds roughly to a combination of the SDSS $i'+z'$ filters \citep{Bessell2005}.

The first block of observations were made over a five-month period, from the end of April 2013 to the end of September 2013. 
Each cluster was observed once or twice a week resulting in between 20 and 40 observations of each cluster. 
Due to the small FoV, it was unfortunately necessary to reject a number of the observations because of bad pointing. 
Because the crowded central parts of the clusters were targeted, we found that it was not possible to gain any useful information from observations that had a seeing of worse than $1\farcs5$ in full-width at half-maximum (FWHM).

Thus only 10 to 15 observations per cluster were found to be useful, which made it nearly impossible to make reasonable period estimates of the variable stars.
It was thus decided to perform a second block of observations over a 6 week period from mid-April 2014 to the start of June 2014. 
This increased the number of epochs to between 28 and 37, which enabled us to derive much better period estimates.

To achieve a reasonable S/N, each observation has a total exposure time of 8 minutes, except for NGC~6441 where they are 10 minutes long. 
With the frame-rate of 10 Hz, the 8 and 10 minute observation consist of 4800 and 6000 single exposures, respectively.

\subsection{EM data reduction}
The algorithms described by \cite{Harpsoe2012} were used to make bias, flat and tip-tilt corrections, and to find the instantaneous image quality (PSF FWHM) for each exposure. 

The tip-tilt correction and image quality are found using the Fourier cross correlation theorem. 
Given a set of $k$ exposures, each containing $i$ pixel columns and $j$ pixel rows, a comparison image $C(i,j)$ is constructed by taking the average of 100 randomly chosen exposures. 
The cross correlation $P_k(i,j)$ between $C$ and a bias- and flat-corrected exposure $I_k(i,j)$ can be found using 
\begin{equation}
P_k(i,j) = \left|FFT^{-1}\left[FFT(C)\cdot \overline{FFT(I_k)}\right]\right|
\end{equation}
where $FFT$ is the fast Fourier transform.
The appropriate shift, that will correct the tip-tilt error, can now be found by locating the $(i,j)$ position of the global maximum in $P_k$.

A measure for the image quality $q_k$ is found by scaling the maximum value of $P_k$ with the sum of its surrounding pixels within a radius $r$ 
\begin{equation}
q_k = \frac{P_k(i_{\text{max}},j_{\text{max}})}{\sum_{\substack{\left|(i-i_{\text{max}},j-j_{\text{max}})\right| < r \\ (i,j)\neq(i_{\text{max}},j_{\text{max}})}} P_k(i,j)}\ .
\end{equation}
Using this factor, instead of just the maximum value of the pixel values in the frame \citep{Smith2009}, helps to mitigate the effects of fluctuations in atmospheric extinction and scintillation that can happen on longer time scales.

A cosmic ray detection and correction algorithm was used on the full set of exposures in each observation. 
For a conventional CCD, where the noise is normally distributed, the mean and sigma values of each $(i,j)$ pixel in the frame over all $k$ exposures can be found. 
Sigma clipping can then be used to reject cosmic rays.
For an EMCCD the noise is not normally distributed, and we have therefore developed a method that takes this into account (Skottfelt et al., in prep.). 
Instead of finding the mean value, the rate of photons is estimated for each pixel in the frame over all the frames. 
Using this photon rate at each pixel position, the probability $p$ for the observed pixel values in each exposure can be calculated and if a pixel value is too improbable, then it is rejected. 

The output of the EM reduction of a single observation is a ten-layer image cube, where each layer is the sum of some percentage of the shift-and-added exposures after the exposures have been organised into ascending order by image quality.
The specific percentages can be modified, but to preserve as much of the spatial information as possible, we have chosen the following (non-cumulative) percentages for the layers: (1,~2,~5,~10,~20,~50,~90,~98,~99,~100).
This means that the first layer contains the sum of the best 1\% of the exposures in terms of image quality, the second layer the second best 1\%, the third the next 3\%, and so on. 
If the shift needed to correct for tip-tilt error is too big (more than 20 pixels) for a given exposure, then the exposure is rejected. 
This is done to ensure that, except for a 20 pixel border, each layer has a uniform number of input exposures.
The total number of exposures used to create the ten-layer image cube might therefore be a little lower than expected.

\subsection{Photometry}

To extract the photometry from the observations, the {\tt DanDIA} pipeline\footnote{{\tt DanDIA} is built from the DanIDL library of IDL routines available at http://www.danidl.co.uk} was used \citep{Bramich2008,Bramich2013}. 

The pipeline has been modified to stack the sharpest layers from all of the available cubes to create a high-resolution reference image, such that an optimal combination of spatial resolution and S/N is achieved. 
Using this method, we were able to achieve resolutions between $0\farcs39$ and $0\farcs53$ for the reference images for the five clusters, and still have a good S/N ratio.
Table~\ref{table:ref_info} reports the combined integration time and FWHM of each of the reference images. 
The peculiar PSFs that can be seen in the reference images (Figs. \ref{fig:NGC6388_ref}, \ref{fig:NGC6441_ref}, \ref{fig:NGC6528_ref}, \ref{fig:NGC6638_ref}, \ref{fig:NGC6652_ref}) come from a triangular coma that is particular to the Danish telescope and this effect manifests itself under good seeing conditions. The telescope was commissioned in 1979 and is therefore not built for such high-resolution instruments. 

\begin{table}
 \caption{Reference image details.}
 \label{table:ref_info}
 \centering
 \begin{tabular}{ccccc}
 \hline\hline
 Cluster & Exp. Time (s) & PSF FWHM & Fig. No.\\
  \hline
NGC~6388  &  240.0 & $0\farcs42$  &\ref{fig:NGC6388_ref} \\
NGC~6441  &  312.0 & $0\farcs39$  &\ref{fig:NGC6441_ref} \\
NGC~6528  &  348.0 & $0\farcs48$  &\ref{fig:NGC6528_ref} \\
NGC~6638  &  374.4 & $0\farcs53$  &\ref{fig:NGC6638_ref} \\
NGC~6652  &  403.2 & $0\farcs46$  &\ref{fig:NGC6652_ref} \\
 \hline
 \end{tabular}
 \tablefoot{Combined exposure time and FWHM 
for the reference images.  The reference images are used to show the positions of the variable stars in Sect. \ref{sec:results} and the last column gives the Figure numbers for these.}
\end{table}

Reference fluxes and positions of the stars are measured from the reference image.
Each ten-layer cube is then stacked into a single science image.
The reference image, convolved with the kernel solution, is subtracted from each of the science images to create difference images,
and, in each difference image, the differential flux for each star is measured by scaling the PSF at the position of the star \citep[for details, see][]{Bramich2011}.

The noise model for EMCCD data differs from that for conventional CCD data \citep{Harpsoe2012}. 
In the EMCCD noise model for a single exposure, the variance in pixel $ij$ is found to be
\begin{equation}
\sigma^2_{ij} = \frac{\sigma^2_0}{F^2_{ij}} + \frac{2\cdot S_{ij}}{F_{ij}\cdot G_{EM} \cdot G_{PA}}
\end{equation}
where $\sigma_0$ is the CCD readout noise (ADU), $F_{ij}$ is the master flat-field image, 
$S_{ij}$ is the image model (sky + object photons), $G_{EM}$ is the EM gain (photons$/e^-$) and $G_{PA}$ is the PreAmp gain ($e^-/$ADU).

When $N$ exposures are combined by summation, the pixel variances of the combined frame are thus
\begin{equation}
\sigma^2_{ij,\text{comb}} = \frac{N \cdot \sigma^2_0}{F^2_{ij}} + \frac{2\cdot S_{ij,\text{comb}}}{F_{ij}\cdot G_{EM} \cdot G_{PA}}.
\end{equation}

The {\tt DanDIA} software has therefore been further modified to employ this noise model.

In Table \ref{table:all_vars}, we outline the format of the data as it is provided at the CDS.

\begin{table*}
 \caption{Format for the time-series photometry of all confirmed variables in our field of view of the five clusters.}
 \label{table:all_vars}
 \centering
 \begin{tabular}{cccccccccccc}
 \hline\hline
 Cluster & \# & Filter & HJD & $M_{\text{std}}$ & $m_{\text{ins}}$ & $\sigma_m$ & $f_{\text{ref}}$ & $\sigma_{\text{ref}}$ & $f_{\text{diff}}$ & $\sigma_{\text{diff}}$ & $p$\\
         &    &      & $(d)$ & (mag)     & (mag)     & (mag)      & (ADU s$^{-1}$) & (ADU s$^{-1}$) & (ADU s$^{-1}$) & (ADU s$^{-1}$)\\        
 \hline
 NGC6388 & V29 & $I$ & 2456436.95140 & 14.620 & 5.617 & 0.003 & 44901.760 & 4397.950 & 54393.674 & 627.525 & 4.6389 \\
 NGC6388 & V29 & $I$ & 2456455.86035 & 14.375 & 5.372 & 0.003 & 44901.760 & 4397.950 & 87182.547 & 578.940 & 3.3446 \\
 \vdots & \vdots & \vdots & \vdots & \vdots & \vdots & \vdots & \vdots & \vdots & \vdots & \vdots & \vdots  \\
 NGC6441 & V63  & $I$ & 2456454.78961 & 16.770 & 7.889 & 0.007 & 9899.919 & 2247.418 & -17795.506 & 260.090 & 6.1094 \\
 NGC6441 & V63  & $I$ & 2456509.61751 & 16.561 & 7.680 & 0.005 & 9899.919 & 2247.418 & -8594.633 & 212.423 & 6.0163 \\
 \vdots & \vdots & \vdots & \vdots & \vdots & \vdots & \vdots & \vdots & \vdots & \vdots & \vdots & \vdots  \\
 \hline
 \end{tabular}
 \tablefoot{The standard $M_{\text{std}}$ and instrumental $m_{\text{ins}}$ magnitudes listed in Cols. 5 and 6 respectively correspond to the cluster, variable star, filter and epoch of mid-exposure listed in Cols. 1-4, respectively. 
 The uncertainty on $m_{\text{ins}}$ is listed in Col. 7, which also corresponds to the uncertainty on $M_{\text{std}}$. 
 For completeness, we also list the reference flux $f_{\text{ref}}$ and the differential flux $f_{\text{diff}}$ (Cols. 8 and 10 respectively), along with their uncertainties (Cols. 9 and 11), as well as the photometric scale factor $p$. 
 Instrumental magnitudes are related to the other quantities via $m_{\text{ins}} = 17.5 - 2.5 \cdot \text{log}_{10}(f_{\text{ref}} + f_{\text{diff}}/p)$.
 This is a representative extract from the full table, which is available at the CDS. 
 }
\end{table*}

\subsection{Astrometry}
HST observations of all five clusters have been performed during different observing campaigns \citep[e.g.][]{Djorgovski1995,Piotto1999,Feltzing2000}. 
We made use of the high spatial resolution and precise astrometry of the HST images to derive an astrometric solution for each of our reference images.

The astrometric calibration was performed using the object detection and field overlay routines in the image display tool GAIA \citep{Draper2000}. 
The astrometric fit was then used to calculate the J2000.0 celestial coordinates for all of the variables in our field of view.

\subsection{Photometric calibration}
Using the colour information that is available from the colour-magnitude diagram (CMD) data (see Sec. \ref{sec:cmd}), rough photometric calibrations can be made for NGC~6388, NGC~6441 and NGC~6652. 

\begin{figure}[t]
   \centering
   \includegraphics[width=\linewidth]{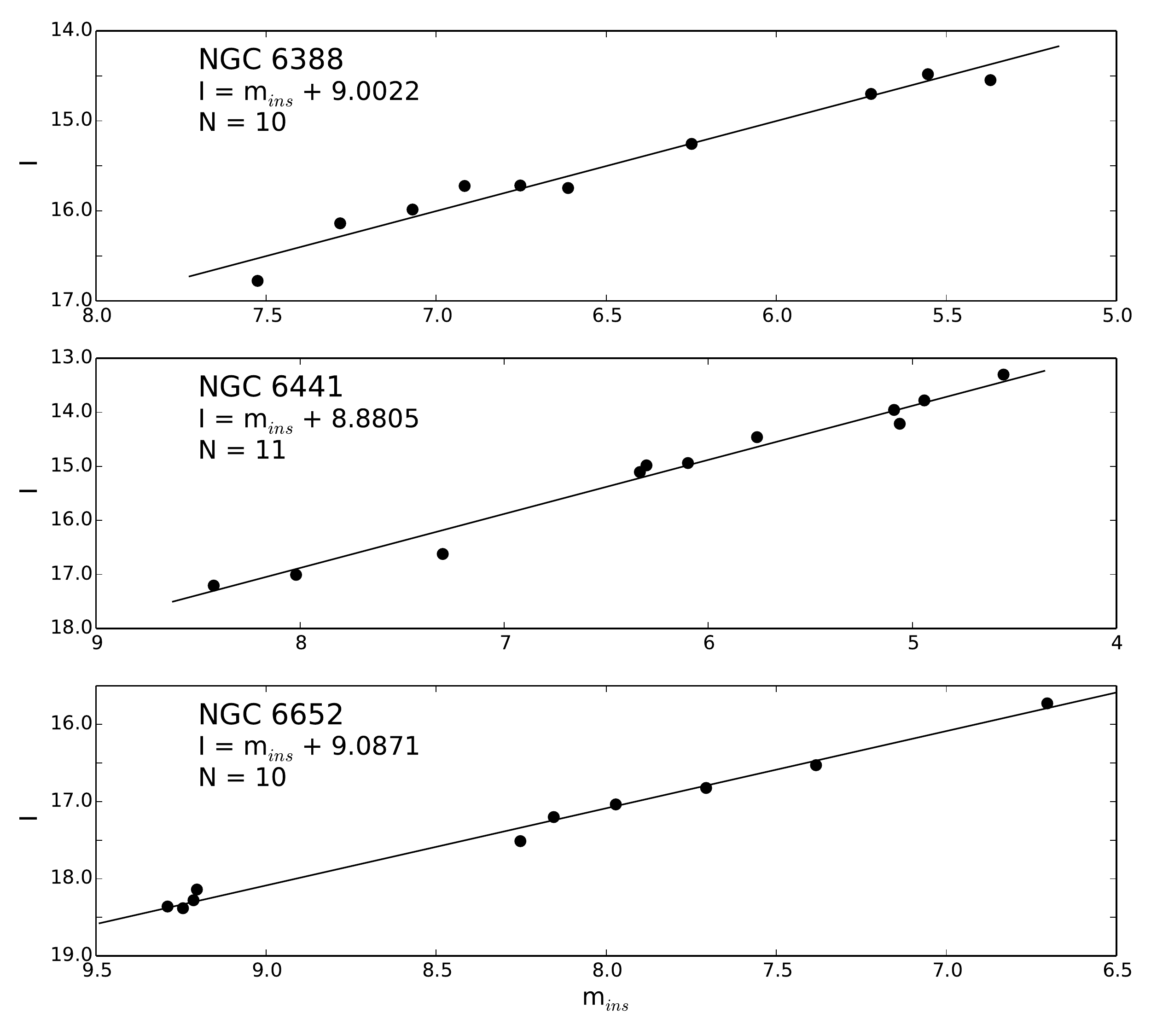}
      \caption{Plot of standard $I$ magnitudes against the instrumental magnitudes. The solid line in each panel shows the best fit calibration.
              }
        \label{fig:photcal}
   \end{figure}

A number of reasonably isolated stars were selected and by matching their positions with the CMD data, their standard $I$ magnitudes were retrieved. 
By finding the offset between the standard $I$ magnitudes and the mean instrumental magnitudes found by {\tt DanDIA}, a rough photometric calibration was found. 
Fig.~\ref{fig:photcal} shows a plot of the standard $I$ magnitudes versus the mean instrumental magnitudes along with the fit for each of the clusters where CMD data are available.

Due to the non-standard filter that is used for these observations, the photometric calibration is only approximate and there is therefore some added uncertainty in the listed $I$ magnitudes.

For NGC~6528 and NGC~6638 where no CMD data are available, we have chosen to adopt the photometric conversion of
\begin{equation}
I = m_{\text{ins}} + 9.0,
\end{equation}
where $m_{\text{ins}}$ is the instrumental magnitude,
to provide approximate $I$ magnitudes for these two clusters as well.

\subsection{Colour-magnitude diagrams} \label{sec:cmd}
As we only have observations in one filter, it is not possible to make a CMD based on our own data. 
Three of the clusters; NGC~6388, NGC~6441 and NGC~6652; are part of \emph{The ACS Survey of Galactic Globular Clusters} \citep{Sarajedini2007}, and 
the data files from the survey, that include $V$ and $I$ magnitudes and celestial coordinates, are available online\footnote{www.astro.ufl.edu/$\sim$ata/public\_hstgc/}.

We were able to match many of the variable stars using celestial coordinates and $I$ magnitudes, enabling us to recover colour information for these objects.
Unfortunately, for some of the variables, their proximity to a bright star meant that we were unable to do this.

For NGC~6528 and NGC~6638, no suitable data to create a CMD were found. The \citet{Piotto2002} study includes NGC6638, but the data files that are available do not contain any celestial coordinates and are therefore not useful for our purposes.


\section{Results} \label{sec:results}
Several methods were used to detect the variable stars in our data. Firstly an image representing the sum of the absolute-valued difference images with pixel values in units of sigma, was constructed for each cluster. 
These images were visually inspected for peaks indicating stars that show signs of variability.
Secondly a diagram of the root-mean-square (RMS) magnitude deviation versus mean magnitude for the calibrated light curves was produced, from which we selected stars with a high RMS for further inspection.
Finally, the difference images were blinked in sequence in order to confirm the variations of all suspected variables.


Period estimates were made using a combination of the string-length statistic $S_Q$ \citep{Dworetsky1983} and the phase dispersion minimisation method \citep{Stellingwerf1978}.

The results for each cluster are presented below. 
Note that the RR Lyrae (RRL) nomenclature introduced by \citet{Alcock2000} has been adopted for this paper; thus RR0 designates an RRL pulsating in the fundamental mode, RR1 designates an RRL pulsating in the first-overtone mode, and RR01 designates a double-mode (fundamental and first-overtone) RRL.

Stars in the upper part of the instability strip are in some parts of the literature referred to as Population II Cepheids (P2C). We have adopted the 'W Virginis' or 'CW' nomenclature, with sub-classifications CWA for CW stars with periods between 8 and 30 days, and CWB for CW stars with periods shorter than 8 days \citep[General Catalog of Variable Stars (GCVS),][]{Samus2009}.
CW stars with periods between 0.8 and 3 days are possibly anomalous Cepheids (AC), which are believed to be too luminous for their periods.
If the period is longer than 30 days, then we classify them as RV Tauri stars (RV)  \citep{Clement2001}.

For variable stars on the red giant branch (RGB), which in some parts of the literature are just referred to as long period variables (LPV), we have adopted the nomenclature of the GCVS.
Thus stars on the RGB showing a noticeable periodicity and with periods over 20 days are classified as semi-regular (SR), and stars that show no evidence of periodicity are classified as long-period irregular (L).
Note that some stars might actually be on the asymptotic giant branch (AGB) and not the RGB. However, distinguishing the two is difficult without a proper spectroscopic analysis and is therefore not possible from the CMDs in this study.


\subsection{NGC~6388}
\subsubsection{Background information} \label{sect:6388_back}
The first 9 variable stars in NGC~6388 were reported by \citet{LloydEvans1973}. These were assigned the numbers V1-V9 by \citet{SawyerHogg1973} in her third catalogue of variable stars in globular clusters.
Three new variables, V10-V12, were found by \citet{LloydEvans1977} using observations from $V$ and $I$ band photographic plates. They also confirmed the existence of the previous 9 variables, but were not able to provide periods for any of the variables.
Using $B$-band photographic plate observations, \citet{Hazen1986} presented 14 new variables, V13-V26, within the tidal radius, and 4 field variables. 

\citet{Silbermann1994} were the first to use CCD observations to search for variable stars in the cluster, and were able to find 3 new variables, V27-V29, and 4 suspected ones. Periods were given for the three new variables and for V17 and V20.

From observations obtained at the 0.9m telescope at the Cerro Tololo Inter-American Observatory (CTIO), \citet{Pritzl2002} (hereafter P02) were able to find 28 new variables, V30-V57, which includes 3 of the 4 suspected variables in \citet{Silbermann1994}. The data for the P02 paper were obtained over 10 days, so only very few long period variables were found. 

Due to the high central concentration (see Table~\ref{table:NGCs}) none of the variable stars found up to this point are located in the central part of the cluster, except for V29.
However, when the P02 data were re-analysed by \citet{Corwin2006} (hereafter C06) using the ISIS version 2.1 image-subtraction package \citep{Alard2000,Alard1998}, 12 new variables, V58-V69, and 6 suspected ones, denoted SV1-SV6, were found. All of these variables are either RRL or CW stars and most of them are located in the central parts of the cluster. 

\subsubsection{This study}

\begin{figure*}[ht]
   \centering
   \includegraphics[width=\linewidth]{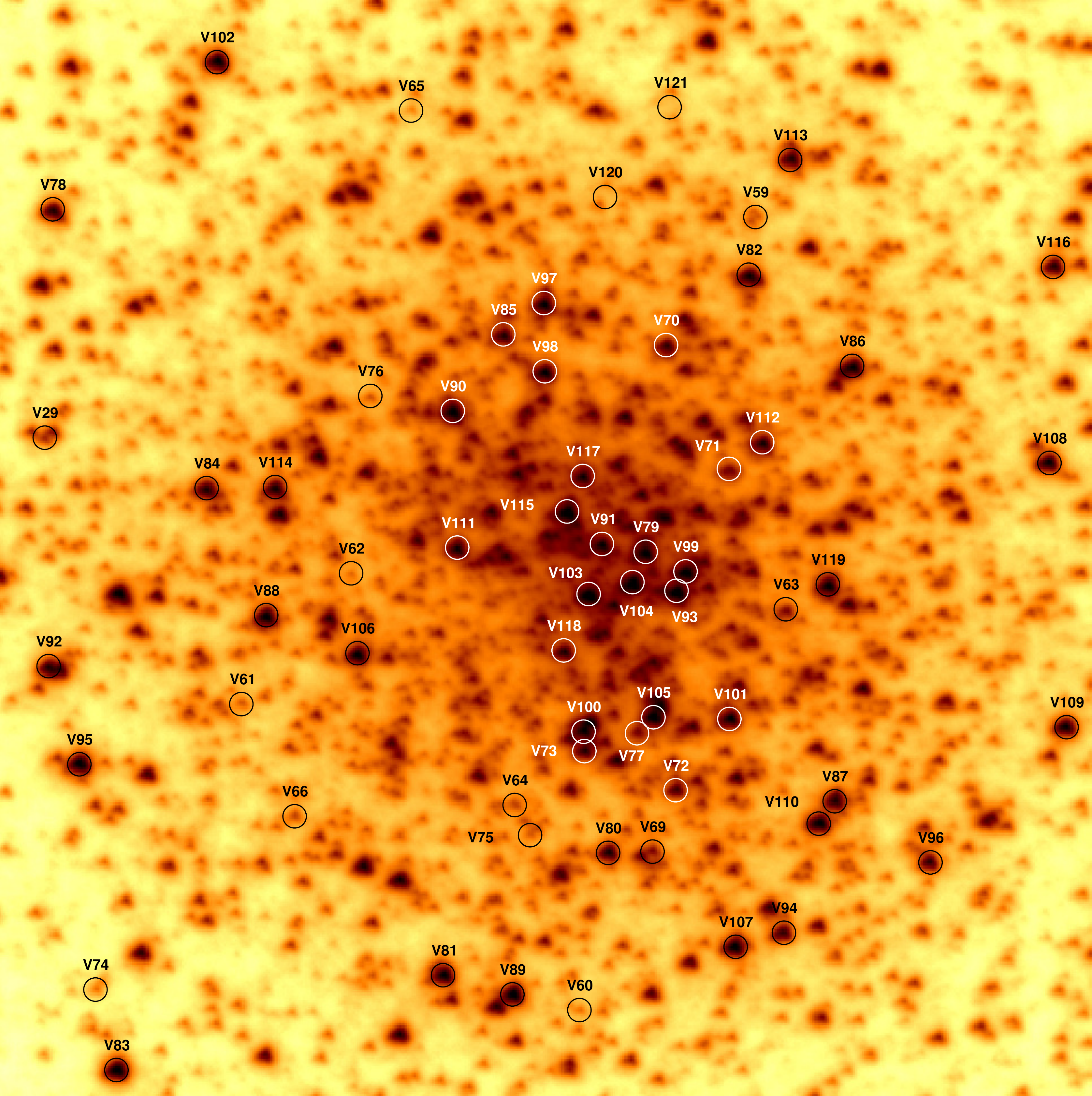}
      \caption{\fchartText{NGC~6388}{Labels and circles are white/black only for clarity. }{41}{41}}
        \label{fig:NGC6388_ref}
   \end{figure*}
   
\begin{figure}[ht]
   \centering
   \includegraphics[width=\linewidth]{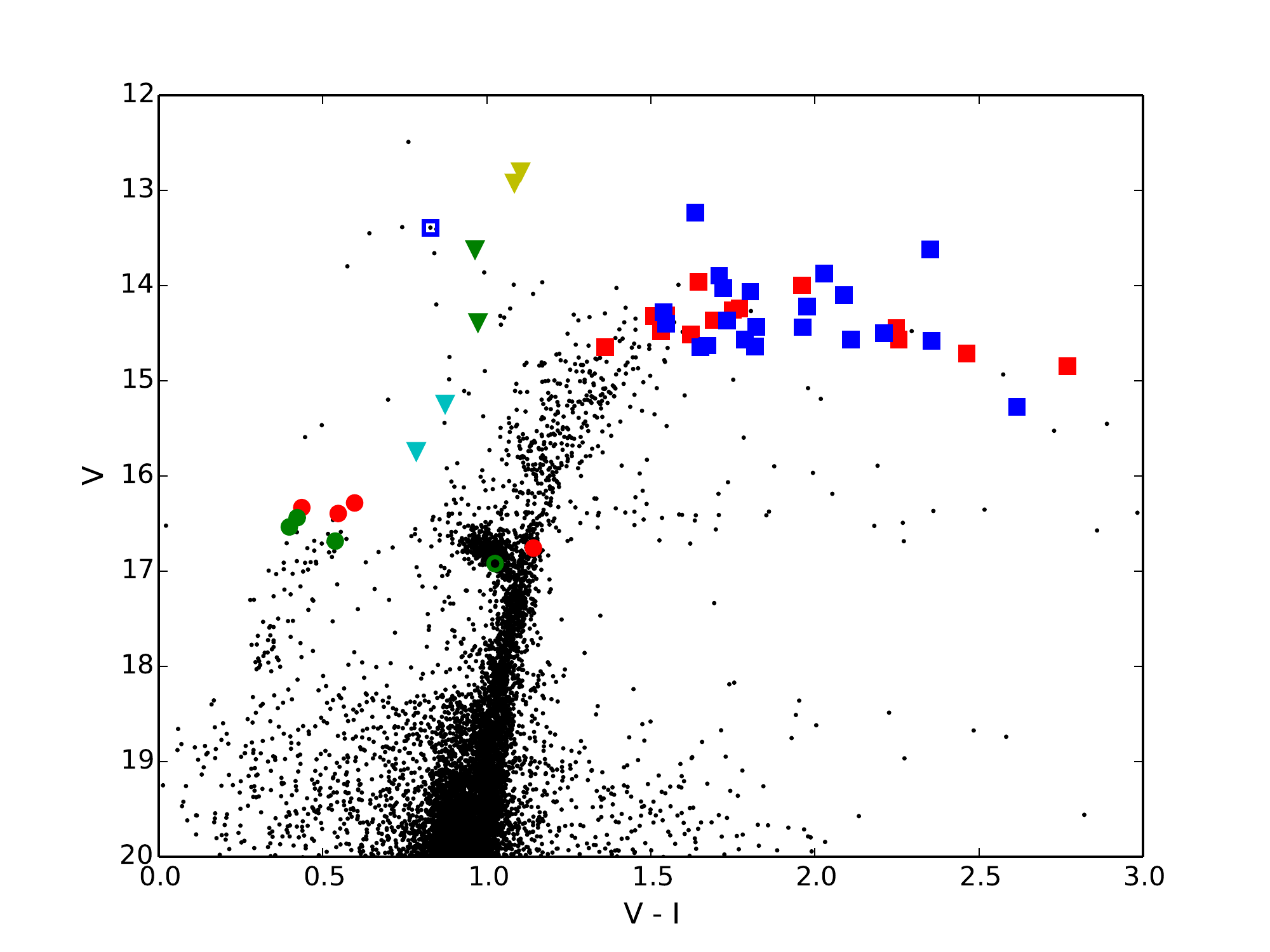}
      \caption{NGC~6388: $(V-I),V$ colour-magnitude diagram made from HST/ACS data as explained in Sect. \ref{sec:cmd}. The stars that show variability in our study are plotted as follows: RRL as filled circles (RR0 in red, RR1 in green, RR01 in blue), CW as filled triangles (CWA in green, CWB in cyan, AC in magenta, RV in yellow), and RGB stars as filled squares (SR in red, L in blue). An open symbol means that the classification is uncertain.
              }
        \label{fig:NGC6388_cmd}
   \end{figure}

\begin{figure}[ht]
   \centering
   \includegraphics[width=\linewidth]{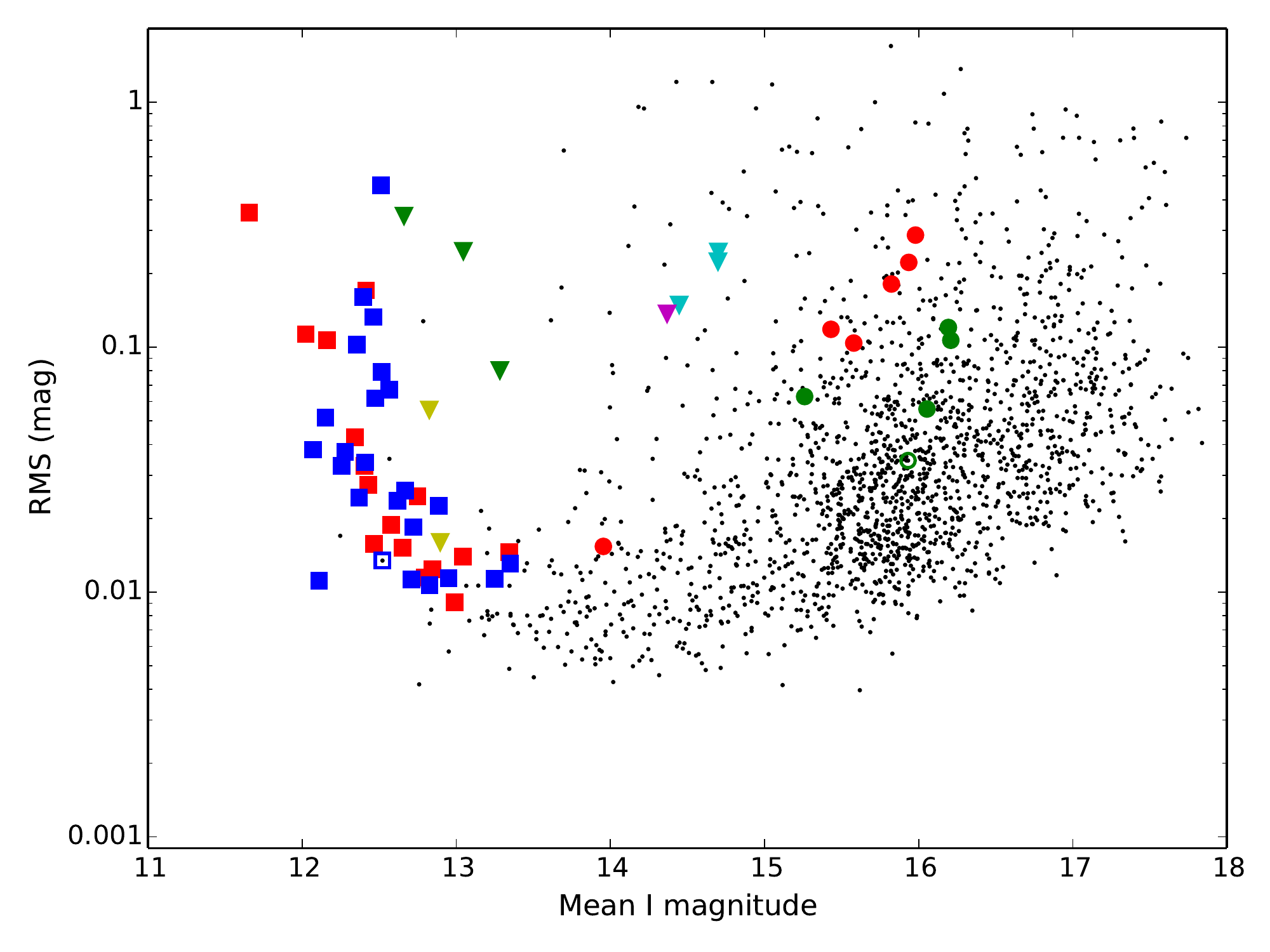}
      \caption{NGC~6388: Plot of the RMS magnitude deviation versus the mean magnitude for each of the 1824 calibrated $I$ light curves. The variables are plotted with the same symbols as in Fig.~\ref{fig:NGC6388_cmd}.
              }
        \label{fig:NGC6388_rms}
   \end{figure}

A finding chart for the cluster containing the variables detected in this study is shown in Fig.~\ref{fig:NGC6388_ref} and Fig.~\ref{fig:NGC6388_cmd} shows a CMD of the stars in our FoV, with the variables overplotted.

Fig.~\ref{fig:NGC6388_rms} shows the RMS magnitude deviation for the 1824 stars with calibrated $I$ light curves versus their mean magnitude. From this plot it is evident that stars fainter than 18th magnitude have not been detected, which can probably be explained by the very dense stellar population in the central region of the cluster. This creates a very high background intensity, which makes it hard to detect faint stars, and which increases the possibility of blending, making some stars appear brighter than they are.

\subsubsection{Known variables}

\begin{table*}
\caption{NGC~6388: Details of the 14 previously known variables in our FoV}
\label{table:NGC6388_V}      
\centering                          
\begin{tabular}{c c c c c c c c c c}        
\hline\hline                 
Var & RA (J2000.0) & Dec. (J2000.0) & Epoch $(d)$ & P $(d)$ & P$_{\mathrm{C06}}$ $(d)$ & $< I >$ & $A_{i'+z'}$ & Blend & Classification\\    
\hline                        
V29 & 17:36:15.321 & -44:44:02.62 & 6772.8616 & 1.8652  & { }1.88\tablefootmark{a} & 14.63 & 0.77 & ii & CWB  \\
V59 & 17:36:17.870 & -44:43:54.77 & 6514.5622 & 0.58884 & 0.589 & 15.48 & 0.39 & ii  & RR0  \\
V60 & 17:36:17.169 & -44:44:24.41 & 6782.7705 & 0.37339 & 0.372 & 16.19 & 0.31 & iii  & RR1  \\
V61 & 17:36:15.995 & -44:44:12.73 & 6514.5622 & 0.65594 & 0.657 & 15.94 & 0.83 & iii & RR0 \\
V62 & 17:36:16.398 & -44:44:07.89 & 6805.8471 & 0.71134 & 0.708 & 15.79 & 0.51 & iii & RR0 \\
V63 & 17:36:17.941 & -44:44:09.50 & 6477.6692 & 2.038   & 2.045 & 14.64 & 0.73 & ii  & CWB \\
V64 & 17:36:16.958 & -44:44:16.68 & 6797.8161 & 0.60137 & 0.595 & 15.35 & 0.43 & ii  & RR0 \\
V65 & 17:36:16.655 & -44:43:50.57 & 6809.7600 & 0.39600 & 0.395 & 16.19 & 0.36 & iii & RR1 \\
V66 & 17:36:16.174 & -44:44:16.97 & 6763.9007 & 0.34989 & 0.350 & 15.25 & 0.19 & ii  & RR1 \\
V69 & 17:36:17.444 & -44:44:18.51 & 6808.8938 & 3.539   & 3.601 & 14.55 & 0.47 & iii & CWB  \\
{ }V70\tablefootmark{b} & 17:36:17.540 & -44:43:59.52 & 6783.8319 & 12.38   & $\sim$8 & 13.28 & 0.24  &  iii & CWA  \\
{ }V71\tablefootmark{c} & 17:36:17.752 & -44:44:04.20 & 6792.8127 & 0.8539  & 0.847   & 13.96 & 0.05  &  i   & RR0  \\
{ }V72\tablefootmark{d} & 17:36:17.532 & -44:44:16.22 & 6784.7647 & 18.01   & $\sim$12 & 13.08 & 0.82 &  iii & CWA  \\
{ }V73\tablefootmark{e} & 17:36:17.211 & -44:44:14.70 & 6792.8127 & 26.1    & $\sim$4.5 & 12.74 & 1.11 & iii   & CWA  \\
\hline                                   
\end{tabular}
\tablefoot{\varTableText{2456522.48}{P$_{\mathrm{C06}}$ are periods from C06, which have been included as a reference.}
\tablefoottext{a}{Period from \citet{Pritzl2002}.}
\tablefoottext{b}{Denoted SV1,}
\tablefoottext{c}{SV2,}
\tablefoottext{d}{SV4, and}
\tablefoottext{e}{SV5 in C06.}}
\end{table*}

\begin{figure*}[h!]
   \centering
   \includegraphics[width=\linewidth]{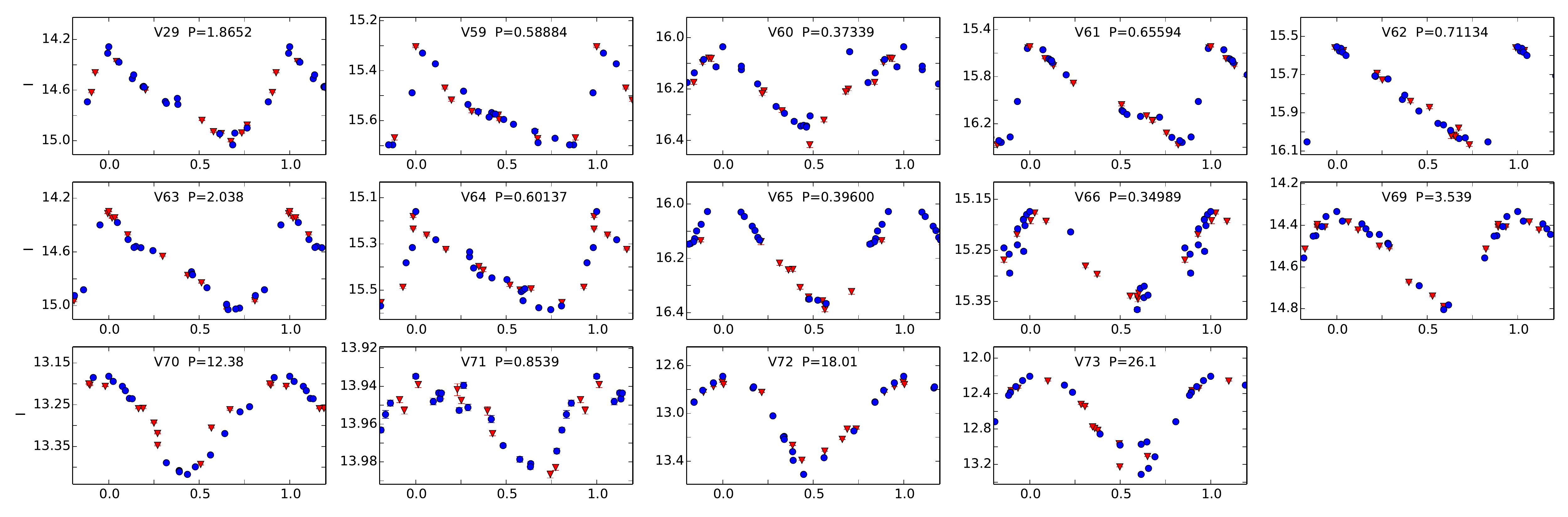}
      \caption{NGC~6388: Phased light curves for the known variables in our FoV. Red triangles are 2013 data and blue circles are 2014 data. Error bars are plotted but are smaller than the data symbols in many cases.}
        \label{fig:NGC6388_V}
   \end{figure*}
   
All of the previously known variables within our FoV are recovered in this analysis.
The light curves for these variables are shown in Fig.~\ref{fig:NGC6388_V} and their celestial coordinates and estimated periods are listed in Table~\ref{table:NGC6388_V}, along with the periods found in C06. The table also reports the mean $I$ magnitude, the amplitude in our filter, and the classification of the variables.

A discussion of individual variables is given below, but generally it can be noted that we find similar periods to those given in C06. Only the three CWs; V70, V72 and V73; have very different periods. In C06 these three stars are listed as suspected variables (SV) due to uncertainty in the classification.

We find some variables to be somewhat brighter than would be expected for their classification. 
This is most likely due to blending with very close neighbours which, as mentioned above, can lead to an overestimation of the reference flux (e.g. see col. 9 of Table \ref{table:NGC6388_V}).
An overestimated reference flux subsequently leads to an underestimation of the amplitude of the variable star.
Errors in the mean magnitudes might also be caused by the rough photometric calibration, but this should only give discrepancies on a much lower level, and should not affect the amplitude.

Unless otherwise stated below, the classifications that are given in P02 or C06 are confirmed by our light curves and the CMD, and have therefore been adopted. A period-luminosity diagram for the stars in the upper part of the instability strip is shown in Fig. \ref{fig:PerLum}, and is discussed in more detail in Sect. \ref{sec:disc6388_6441}.

Discussion of individual variables (square brackets gives approximate position in the CMD as $[V-I,V]$):
\begin{description}
\item[{\bf V29, V63:}]{Based on the relation between period and luminosity compared to the other CW stars, these two stars are classified as CWB stars and not AC stars.
V29 is the only variable that is also found in the P02 paper, i.e. where difference image analysis was not used.  }
\item[{\bf V59, V60:}]{No matching stars have been found in the CMD for these two variables. However, their periods, mean magnitude, and amplitude strongly indicate that the classifications as RR0 and RR1, respectively, are correct.}
\item[{\bf V61-V62, V64-V65:}]{CMD positions, periods, magnitudes and amplitudes all verify the classifications given in C06.}
\item[{\bf V66:}]{This RR1 seems to have an overestimated mean magnitude, and correspondingly underestimated amplitude. This is most likely caused by blending with another star, as the position in the CMD $[0.4,16.5]$ verifies the classification. }
\item[{\bf V69:}]{No CMD position for this star is found, but based on its period and mean magnitude it is classified as a CWB. }
\item[{\bf V70, V73:}]{Those are two of the suspected variables from C06 that are also in our FoV. We report reliable periods in Table \ref{table:NGC6388_V}. The periods and CMD positions ($[1.0,14.3],[1.0,13.6]$) indicate that they are CWA stars. }
\item[{\bf V71:}]{This star is also a suspected variable from C06. The star is highly blended with a brighter star which leads to a mean magnitude that is too bright and a heavily underestimated amplitude. No CMD position is available, but its period and light curve strongly suggest that the star is an RR0.}
\item[{\bf V72:}]{This is the last of the C06 suspected variables in our FoV, and we report a reliable period in Table \ref{table:NGC6388_V}. No CMD position is found for this variable, so it could be a SR, but CWA is a more likely classification considering the relationship between magnitude and period compared to the other CW stars (see Fig. \ref{fig:PerLum})}.
\end{description}

\subsubsection{New variables}

\begin{table*}
\caption{NGC~6388: Details of the 48 new variables found in the cluster. }
\label{table:NGC6388_NV}      
\centering                          
\begin{tabular}{c c c c c c c c c}        
\hline\hline                 
Var & RA (J2000.0) & Dec. (J2000.0) & Epoch $(d)$ & P $(d)$ & $< I >$ & $A_{i'+z'}$ & Blend & Classification\\    
\hline                        
V74 & 17:36:15.449 & -44:44:23.35 & 6790.7919 & 0.35840 & 15.96 & 0.12 & ii  & RR1?  \\
V75 & 17:36:17.010 & -44:44:17.82 & 6782.7705 & 0.40393 & 16.07 & 0.22 & ii  & RR1  \\
V76 & 17:36:16.483 & -44:44:01.25 & 6792.8127 & 0.7574 & 15.98 & 0.67 &  iii & RR0  \\
V77 & 17:36:17.400 & -44:44:14.06 & 6805.8471 & 1.8643 & 14.30 & 0.44 &  iii & AC  \\
V78 & 17:36:15.371 & -44:43:54.06 & 6522.4888 & 24.0 & 12.80 & 0.04 &    iii & SR  \\
V79 & 17:36:17.448 & -44:44:07.26 & 6522.4888 & 27.07 & 12.46 & 0.05 &   ii  & SR  \\
V80 & 17:36:17.286 & -44:44:18.53 & 6488.7479 & 30.28 & 12.86 & 0.20 &   iii & RV \\
V81 & 17:36:16.687 & -44:44:23.02 & 6488.7479 & 33.78 & 12.58 & 0.08 &   iii & SR  \\
V82 & 17:36:17.841 & -44:43:56.93 & 6476.6758 & 40.47 & 12.89 & 0.06 &   ii  & RV \\
V83 & 17:36:15.516 & -44:44:26.38 & 6460.6572 & 50.68 & 12.42 & 0.15 &   iii & SR  \\
V84 & 17:36:15.892 & -44:44:04.61 & 6805.8471 & 55.5 & 12.85 & 0.05 &    iii & SR  \\
V85 & 17:36:16.962 & -44:43:59.02 & 6544.5238 & 56.8 & 13.03 & 0.07 &    iii & SR  \\
V86 & 17:36:18.200 & -44:44:00.41 & 6772.8616 & 57.0 & 12.75 & 0.08 &    ii  & SR  \\
V87 & 17:36:18.097 & -44:44:16.73 & 6518.4918 & 58.3 & 12.99 & 0.04 &    ii  & SR  \\
V88 & 17:36:16.092 & -44:44:09.41 & 6790.7919 & 58.3 & 12.20 & 0.35 &    iii & SR  \\
V89 & 17:36:16.932 & -44:44:23.78 & 6455.8603 & 70.7 & 12.42 & 0.11 &    iii & SR  \\
V90 & 17:36:16.775 & -44:44:01.86 & 6776.8346 & 70.8 & 12.34 & 0.14 &    ii  & SR  \\
V91 & 17:36:17.293 & -44:44:06.94 & 6808.8938 & 113.9 & 12.65 & 0.05 &   iii & SR  \\
V92 & 17:36:15.313 & -44:44:11.19 & 6436.9514 & 114.2 & 12.60 & 0.65 &   ii  & SR  \\
V93 & 17:36:17.554 & -44:44:08.74 & 6436.9514 & 157 & 11.97 & 0.43 &     iii & SR  \\
V94 & 17:36:17.904 & -44:44:21.63 & 6763.9007 & 177 & 13.34 & 0.06 &     iii & SR  \\
V95 & 17:36:15.413 & -44:44:14.89 & 6776.8346 & 180 & 11.91 & 1.06 &     iii & SR  \\
V96 & 17:36:18.432 & -44:44:19.08 &  -  &  -  & 12.51 & 1.55 &   iii & L  \\
V97 & 17:36:17.109 & -44:43:57.87 &  -  &  -  & 12.43 & 0.84 &   iii & L  \\
V98 & 17:36:17.106 & -44:44:00.43 &  -  &  -  & 12.52 & 0.60 &   iii & L  \\
V99 & 17:36:17.588 & -44:44:08.02 &  -  &  -  & 12.37 & 0.35 &   iii & L  \\
V100 & 17:36:17.210 & -44:44:13.97 &  -  &  -  & 12.53 & 0.34 &  ii  & L  \\
V101 & 17:36:17.730 & -44:44:13.58 &  -  &  -  & 12.58 & 0.26 &  iii & L  \\
V102 & 17:36:15.969 & -44:43:48.64 &  -  &  -  & 12.54 & 0.21 &  iii & L  \\
V103 & 17:36:17.241 & -44:44:08.81 &  -  &  -  & 12.13 & 0.20 &  iii & L  \\
V104 & 17:36:17.398 & -44:44:08.39 &  -  &  -  & 12.07 & 0.15 &  iii & L  \\
V105 & 17:36:17.460 & -44:44:13.47 &  -  &  -  & 12.27 & 0.14 &  ii  & L  \\
V106 & 17:36:16.413 & -44:44:10.89 &  -  &  -  & 12.28 & 0.12 &  iii & L  \\
V107 & 17:36:17.731 & -44:44:22.13 &  -  &  -  & 12.41 & 0.12 &  iii & L  \\
V108 & 17:36:18.893 & -44:44:04.17 &  -  &  -  & 12.68 & 0.12 &  iii & L  \\
V109 & 17:36:18.929 & -44:44:14.09 &  -  &  -  & 12.63 & 0.09 &  iii & L  \\
V110 & 17:36:18.038 & -44:44:17.56 &  -  &  -  & 12.38 & 0.08 &  iii & L  \\
V111 & 17:36:16.778 & -44:44:06.99 &  -  &  -  & 12.73 & 0.08 &  iii & L  \\
V112 & 17:36:17.873 & -44:44:03.22 &  -  &  -  & 12.88 & 0.07 &  iii & L  \\
V113 & 17:36:17.999 & -44:43:52.65 &  -  &  -  & 13.35 & 0.06 &  iii & L  \\
V114 & 17:36:16.136 & -44:44:04.61 &  -  &  -  & 12.71 & 0.05 &  iii & L  \\
V115 & 17:36:17.172 & -44:44:05.70 &  -  &  -  & 12.11 & 0.04 &  ii  & L  \\
V116 & 17:36:18.924 & -44:43:56.82 &  -  &  -  & 13.24 & 0.04 &  iii & L  \\
V117 & 17:36:17.231 & -44:44:04.38 &  -  &  -  & 12.83 & 0.04 &  iii & L  \\
V118 & 17:36:17.147 & -44:44:10.91 &  -  &  -  & 12.95 & 0.04 &  iii & L  \\
V119 & 17:36:18.093 & -44:44:08.59 &  -  &  -  & 12.53 & 0.05 &  iii & L?  \\
V120 & 17:36:17.337 & -44:43:53.93 &  -  &  -  &  -  &  -  &     -   & ?  \\
V121 & 17:36:17.575 & -44:43:50.60 &  -  &  -  &  -  &  -  &     -   & ?  \\
\hline                                   
\end{tabular}
\tablefoot{\varTableText{2456522.48}{}}
\end{table*}

\begin{figure*}[ht]
   \centering
   \includegraphics[width=\linewidth]{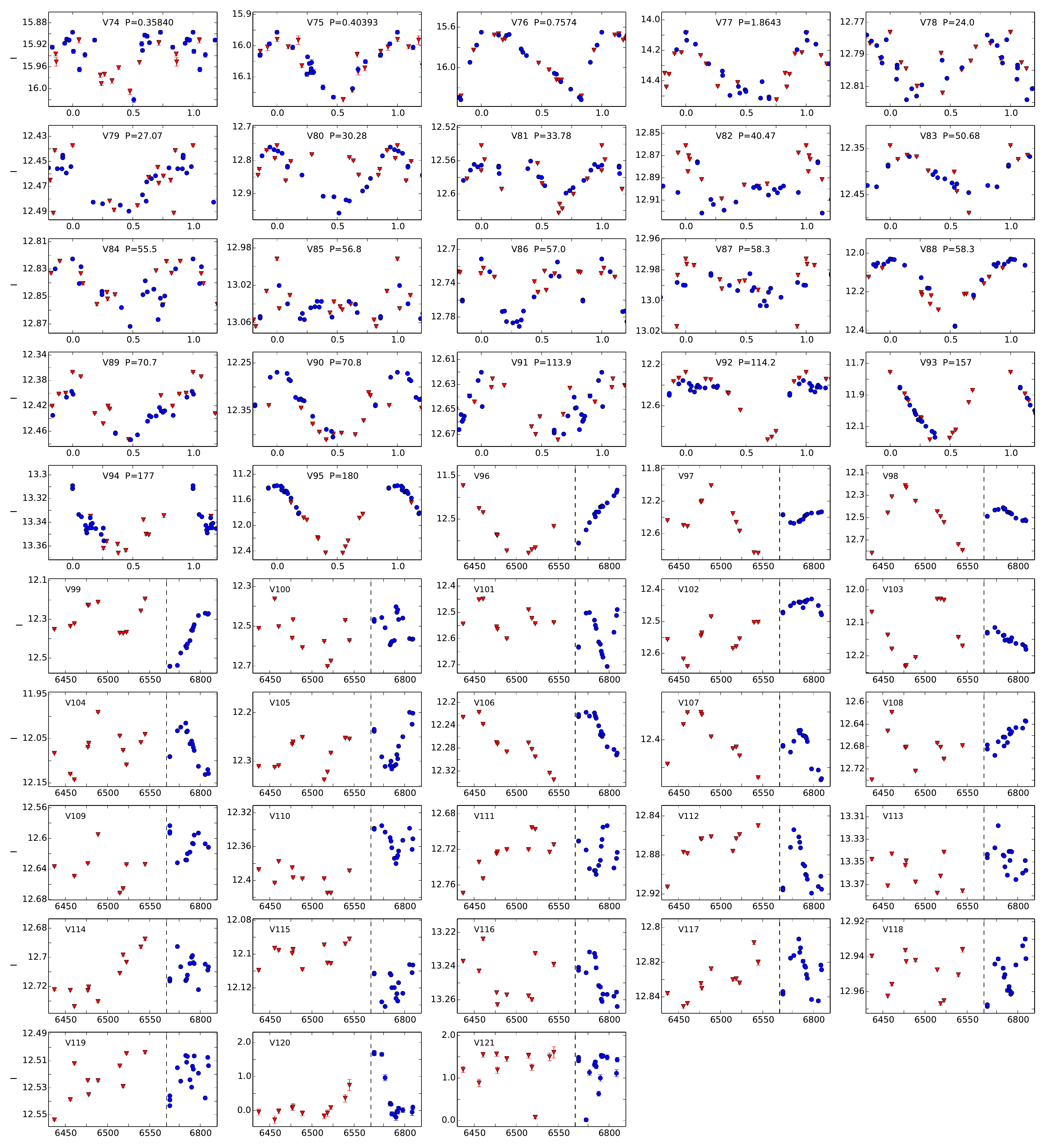}
      \caption{\VarFigText{NGC~6388}{the new} Note that V120 and V121 are plotted in differential flux units, $10^3$ ADU/s, and not calibrated $I$ magnitudes.}
        \label{fig:NGC6388_NV}
   \end{figure*}

In this study we were able to find 48 new variable stars for NGC~6388. Their light curves are shown in Fig.~\ref{fig:NGC6388_NV} and their details are listed in Table~\ref{table:NGC6388_NV}.

Most of the new variables are RGB stars and many of them have small amplitudes, which is probably the reason why they have not been detected until now. 
There are however also two (possibly three) previously undetected RRL stars and three new CW stars. 
For many of the new long period variables it has been hard to determine a period, which may indicate that they are irregular.

Discussion of individual variables:
\begin{description}
\item[{\bf V74:}]{This star has a very noisy light curve, where a period matching an RR1 has been estimated. The mean magnitude also matches this classification, but it falls on the red-clump in the CMD $[1.0,16.9]$. Our RR1 classification is therefore tentative.}
\item[{\bf V75:}]{The light curve and the position of the star in the CMD $[0.5,16.7]$, strongly suggest that this is an RR1.}
\item[{\bf V76:}]{This star has a very good light curve, where the light curve shape, period and mean magnitude strongly indicate an RR0. However, the CMD position $[1.1,16.8]$ is a little discrepant.}
\item[{\bf V77:}]{No CMD information is available for this star, but as the luminosity seems a bit too high compared to its period, we classify it as a possible AC (see Fig. \ref{fig:PerLum}). The star is located very close to another and much brighter variable V105, and it would have been very difficult to resolve these stars using conventional imaging.}
\item[{\bf V78, V79, V81:}]{These three stars have similar positions in the CMD $[1.7,14]$, and  they are therefore most likely SR stars.}
\item[{\bf V80,V82:}]{These two stars are most likely RV stars, based on their position in the CMD $[1.1,12.8]$ and their relation between period and luminosity (see Fig. \ref{fig:PerLum}).}
\item[{\bf V83-V92,V94,V95:}]{All of these stars are on the RGB and combined with their long periods, these can be classified as SR stars. Most of the stars have fairly small amplitudes (0.04~--~0.15~mag)}
\item[{\bf V92, V93:}]{These two stars are not in the CMD, but based on their periods and luminosity they are most likely SR stars.}
\item[{\bf V96-V118}:]{These stars are also on the RGB, but it has not been possible to find any periods that phase their light curves satisfactorily. We have therefore classified them as L stars.}
\item[{\bf V119:}]{The position of this star in the CMD $[0.8,13.4]$ puts it in the CW region. However, we have not been able to find a period that phases the light curve properly and the star has therefore been tentatively classified as an L star.}
\item[{\bf V120, V121:}]{These two stars were not identified by the pipeline and no stars are visible at these positions on the reference image. However, when blinking the difference images some variability is clearly seen and the differential fluxes have thus been measured for these positions in the difference images.
This means that there is no measurement of the reference flux and magnitudes or amplitudes are therefore not given. 
The fact that the stars are not visible on the reference image is probably because they are very faint at the epoch of the reference image.
No good candidates for the stars have been found in the CMD, and no reasonable periods could be estimated.
Without more information it is hard to classify the two stars, but one possibility could be a type of cataclysmic variable, as these are known to have prolonged low and high states, which can be quasi-periodic or have no clear periodicity.}
\end{description}

\subsection{NGC~6441}
\subsubsection{Background information}
The first 10 variables in NGC~6441 were found by \citet{Fourcade1964}.
In \citet{Hesser1976} the authors report that they may have found two variable stars. 
These two stars are, however, found to be non-variable in \citet{Layden1999}, whom made use of CCD imaging. 
\citet{Layden1999} were able to identify, classify, and determine periods for 31 long period variables, 11 RRL stars and 4 eclipsing binaries. A further 9 suspected variables were found but not classified.
From CTIO observations, \citet{Pritzl2001} were able to find 48 new variables, of which 35 are RRL stars. 

Similar to NGC~6388, this cluster has a high central concentration (see Table~\ref{table:NGCs}), and none of the variable stars found up to this point are located in the central part of the cluster, except for V63. \citet{Pritzl2003} (hereafter P03) used HST observations to reveal 41 previously undiscovered variables, V105-V145, the main part of which are located in the central parts.

The \citet{Pritzl2001} data were also re-analysed using ISIS in the C06 paper. In this analysis, five new variables, V146-V150, were found (and recovered in the HST data), but three variables (V136, V138, and V145) that were found in the HST data were not found in the ISIS analysis - all three bona-fide variables located within $10^{\prime\prime}$ of the centre of the cluster. 

\subsubsection{This study}

\begin{figure*}[ht]
   \centering
   \includegraphics[width=\linewidth]{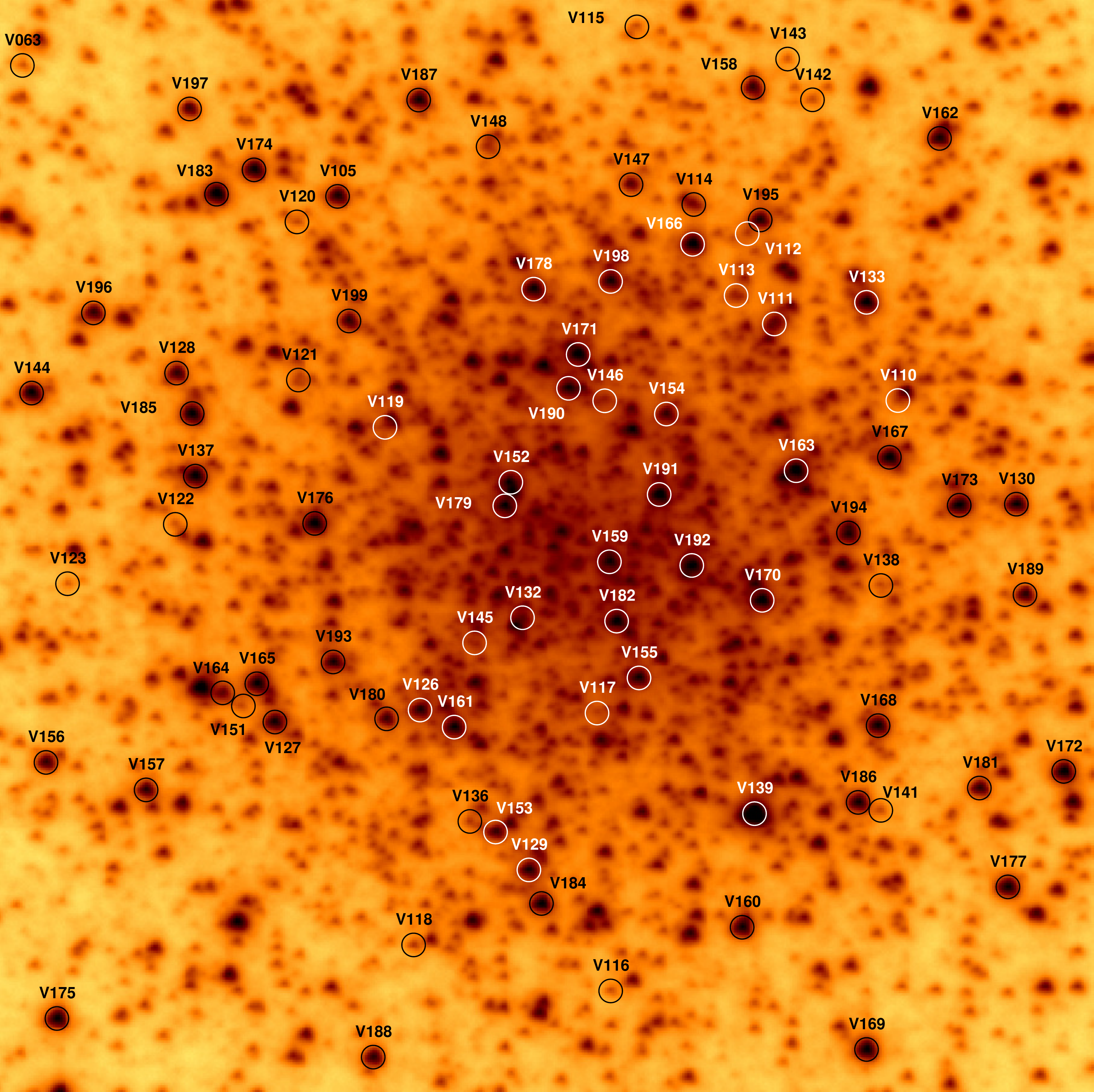}
      \caption{\fchartText{NGC~6441}{Labels and circles are white/black only for clarity. }{41}{41}}
        \label{fig:NGC6441_ref}
   \end{figure*}

\begin{figure}[ht]
   \centering
   \includegraphics[width=\linewidth]{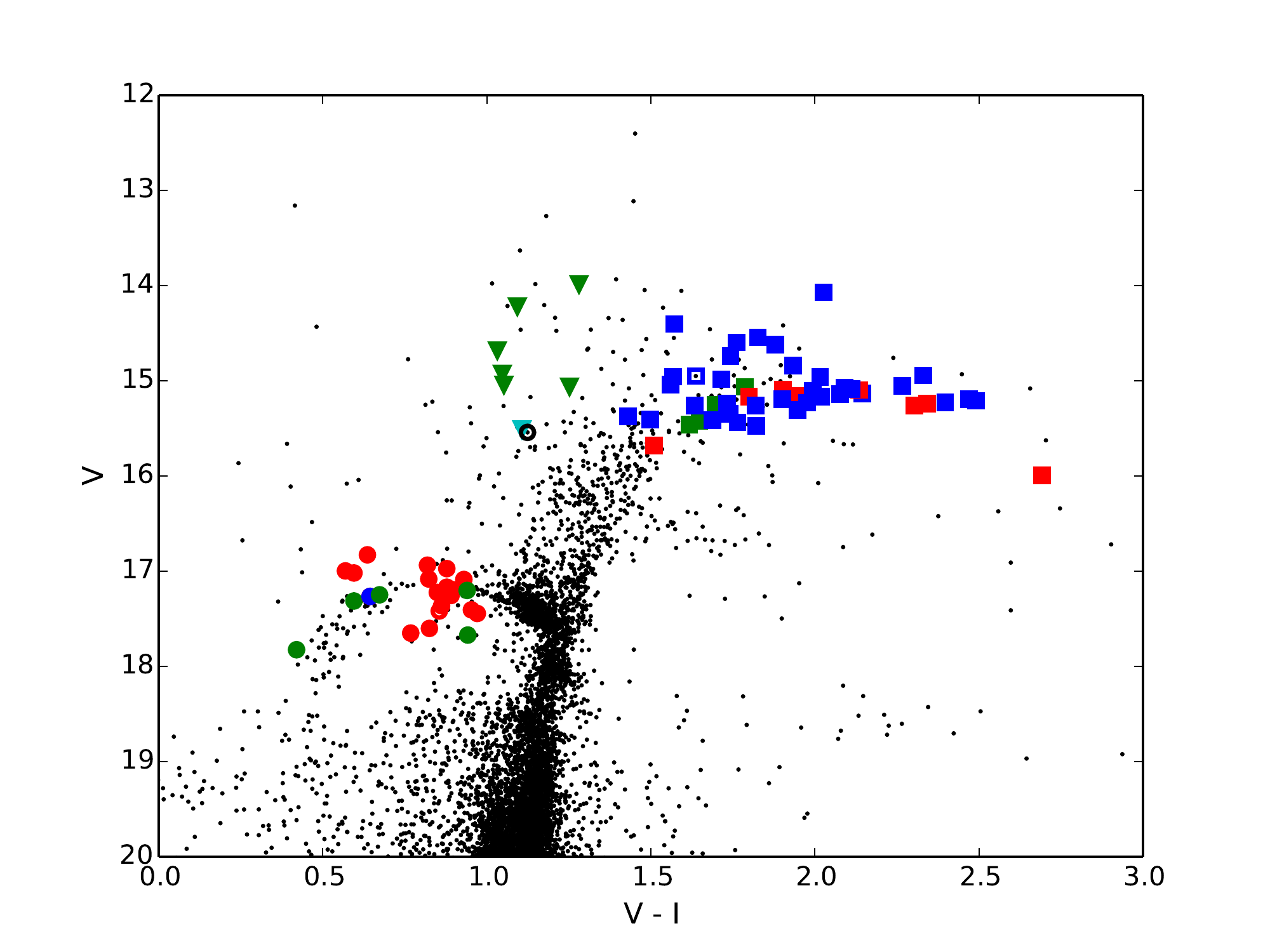}
      \caption{NGC~6441: $(V-I),V$ colour-magnitude diagram made from HST/ACS data as explained in Sect. \ref{sec:cmd}. The stars that show variability in our study are plotted with the same symbols as in Fig. \ref{fig:NGC6388_cmd}.
              }
        \label{fig:NGC6441_cmd}
   \end{figure}

\begin{figure}[ht]
   \centering
   \includegraphics[width=\linewidth]{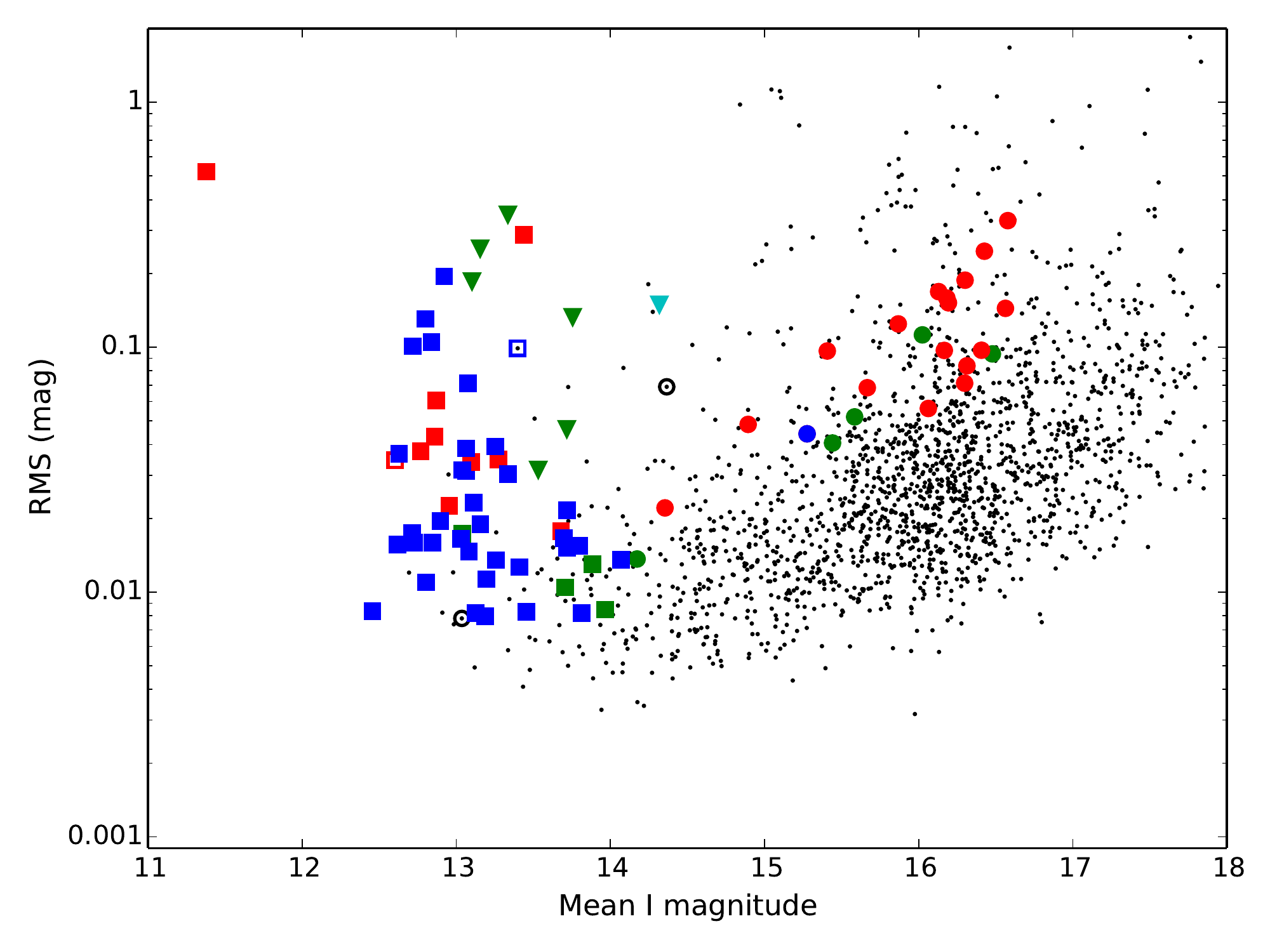}
      \caption{NGC~6441: Plot of the RMS magnitude deviation versus the mean magnitude for each of the 1860 calibrated $I$ light curves. The variables are plotted with the same symbols as in Fig.~\ref{fig:NGC6388_cmd}.
              }
        \label{fig:NGC6441_rms}
   \end{figure}

A finding chart for the cluster containing the variables detected in this study is shown in Fig.~\ref{fig:NGC6441_ref}, and Fig.~\ref{fig:NGC6441_cmd} shows a CMD of the stars within our FoV, with the variables overplotted.

Fig.~\ref{fig:NGC6441_rms} shows the RMS magnitude deviation for the 1860 stars with calibrated $I$ magnitudes versus their mean magnitude. From this plot it is evident that stars fainter than 18th magnitude have not been detected, which can probably be explained by the very dense stellar population in the central region, similar to NGC~6388.

\subsubsection{Known variables}

\begin{table*}[ht]
\caption{NGC~6441: Details of the 35 previously known variables within our FoV.}
\label{table:NGC6441_V}      
\centering                          
\begin{tabular}{c c c c c c c c c c c}        
\hline\hline                 
Var & RA (J2000.0) & Dec. (J2000.0) & Epoch $(d)$ & P $(d)$ & P$_{\mathrm{P03}}$ $(d)$ & P$_{\mathrm{C06}}$ $(d)$ & $< I >$ & $A_{i'+z'}$ & Blend & Classification\\    
\hline                        
V63 & 17:50:11.338 & -37:02:47.42 & 6541.5439 & 0.69789 & 0.69781 & { }0.700\tablefootmark{a} & 16.60 & 0.45 & iii & RR0  \\
V105 & 17:50:12.320 & -37:02:52.61 & 6476.7769 & 113.6   & 111.6   & -     & 13.11 & 0.88 & iii & SR \\
V110 & 17:50:14.068 & -37:03:00.76 & 6770.9142 & 0.76869 & 0.76867 & 0.769 & 16.23 & 0.58 & i   & RR0  \\
V111 & 17:50:13.685 & -37:02:57.78 & 6509.6175 & 0.74464 & 0.74464 & 0.743 & 14.35 & 0.09 & i   & RR0  \\
V112 & 17:50:13.607 & -37:02:54.37 & 6454.7896 & 0.61415 & 0.61419 & 0.614 & 16.23 & 0.78 & iii & RR0 \\
V113 & 17:50:13.567 & -37:02:56.68 & 6541.5439 & 0.58846 & 0.58845 & 0.586 & 15.40 & 0.36 & ii  & RR0  \\
V114 & 17:50:13.441 & -37:02:53.23 & 6541.5439 & 0.67389 & 0.67389 & 0.675 & 14.88 & 0.18 & i   & RR0  \\
V115 & 17:50:13.276 & -37:02:46.52 & 6795.7897 & 0.86315 & 0.86311 & 0.860 & 16.22 & 0.39 & ii  & RR0 \\
V116 & 17:50:13.117 & -37:03:22.64 & 6476.7769 & 0.58229 & 0.58229 & 0.582 & 16.43 & 0.89 & iii & RR0 \\
V117 & 17:50:13.096 & -37:03:12.21 & 6781.7949 & 0.74537 & 0.74529 & 0.745 &  -    &  -   & ii  & RR0  \\
V118 & 17:50:12.500 & -37:03:20.74 & 6805.8572 & 0.9792  & 0.97923 & 0.979 & 16.09 & 0.54 & iii & RR0  \\
V119 & 17:50:12.451 & -37:03:01.31 & 6541.5439 & 0.68627 & 0.68628 & 0.686 &  -    &  -   & i   & RR0  \\
V120 & 17:50:12.190 & -37:02:53.53 & 6789.7942 & 0.36396 & 0.36396 & 0.364 & 16.06 & 0.42 & ii  & RR1  \\
V121 & 17:50:12.182 & -37:02:59.46 & 6773.8400 & 0.83748 & 0.83748 & 0.848 & 15.66 & 0.25 & ii  & RR0 \\
V122 & 17:50:11.783 & -37:03:04.76 & 6789.7942 & 0.74270 & 0.74270 & 0.744 & 16.17 & 0.47 & iii & RR0  \\
V123 & 17:50:11.439 & -37:03:06.89 & 6792.8040 & 0.33566 & 0.33566 & 0.336 & 16.49 & 0.28 & iii & RR1 \\
V126 & 17:50:12.539 & -37:03:11.94 & 6782.7959 & 20.62   & 20.625  & -     & 13.28 & 1.04 & iii & CWA \\
V127 & 17:50:12.081 & -37:03:12.26 & 6775.8182 & 19.77   & 19.773  & -     & 13.27 & 0.74 & iii & CWA \\
V128 & 17:50:11.799 & -37:02:59.09 & 6790.8008 & 13.519  & 13.519  & -     & 13.75 & 0.41 & iii & CWA \\
V129 & 17:50:12.869 & -37:03:18.03 & 6564.5424 & 17.83   & 17.832  & -     & 13.18 & 0.59 & iii & CWA \\
V130 & 17:50:14.433 & -37:03:04.75 & 6476.7769 & 58.00   & 48.90   & -     & 13.39 & 0.38 & iii & SR?, L? \\
V132 & 17:50:12.869 & -37:03:08.57 & 6784.7970 & 2.547   & 2.54737 & -     & 14.32 & 0.44 & i   & CWB \\
V133 & 17:50:13.977 & -37:02:57.05 &  -        &  -      & 122.9   & -     & 12.85 & 0.41 & iii & L \\
V136 & 17:50:12.687 & -37:03:16.16 & 6791.7776 & 0.80574 & 0.80573 & -     & 15.87 & 0.47 & ii  & RR0 \\
V137 & 17:50:11.850 & -37:03:02.97 &  -        &  -      & 51.2    & -     & 13.12 & 0.12 & iii & L \\
V138 & 17:50:14.000 & -37:03:07.68 & 6797.8236 & 0.8020  & 0.80199 & -     & 16.34 & 0.27 & iii & RR0 \\
V139 & 17:50:13.584 & -37:03:16.12 & 6541.5439 & 249     & 249.1   & -     & 12.20 & 1.72 & iii & SR \\
V141 & 17:50:13.982 & -37:03:16.11 & 6789.7942 & 0.8446  & 0.84475 & 0.847 & 16.08 & 0.20 & iii & RR0 \\
V142 & 17:50:13.823 & -37:02:49.41 & 6784.7970 & 0.8840  & 0.88400 & 0.887 & 16.28 & 0.25 & iii & RR0 \\
V143 & 17:50:13.748 & -37:02:47.86 & 6781.7949 & 0.8628  & 0.86279 & 0.863 & 16.42 & 0.32 & iii & RR0 \\
V144 & 17:50:11.341 & -37:02:59.71 & 6782.7959 & 70.6    & 70.6    & -     & 13.09 & 0.11 & iii & SR?, L? \\
V145 & 17:50:12.716 & -37:03:09.47 & 6770.8270 & { }0.55588\tablefootmark{b} & 0.55581 & -     & 15.29 & 0.16 & ii  & RR01  \\
V146 & 17:50:13.145 & -37:03:00.51 & 6797.8236 & 0.40232 & -       & 0.402 & 15.57 & 0.18 & ii  & RR1  \\
V147 & 17:50:13.245 & -37:02:52.43 & 6455.7466 & 0.35487 & -       & 0.355 & 14.17 & 0.05 & i   & RR1  \\
V148 & 17:50:12.798 & -37:02:50.88 & 6454.7896 & 0.39045 & -       & 0.390 & 15.46 & 0.17 & ii  & RR1 \\
\hline                                   
\end{tabular}
\tablefoot{\varTableText{2456541.54}{P$_{\mathrm{P03}}$ and P$_{\mathrm{C06}}$, are periods from P03 and C06, respectively, which have been included as a reference.}
\tablefoottext{a}{Period from \citet{Pritzl2001}.}
\tablefoottext{b}{First-overtone period; fundamental period is found to be 0.72082 d.}}
\end{table*}

\begin{figure*}[ht]
   \centering
   \includegraphics[width=\linewidth]{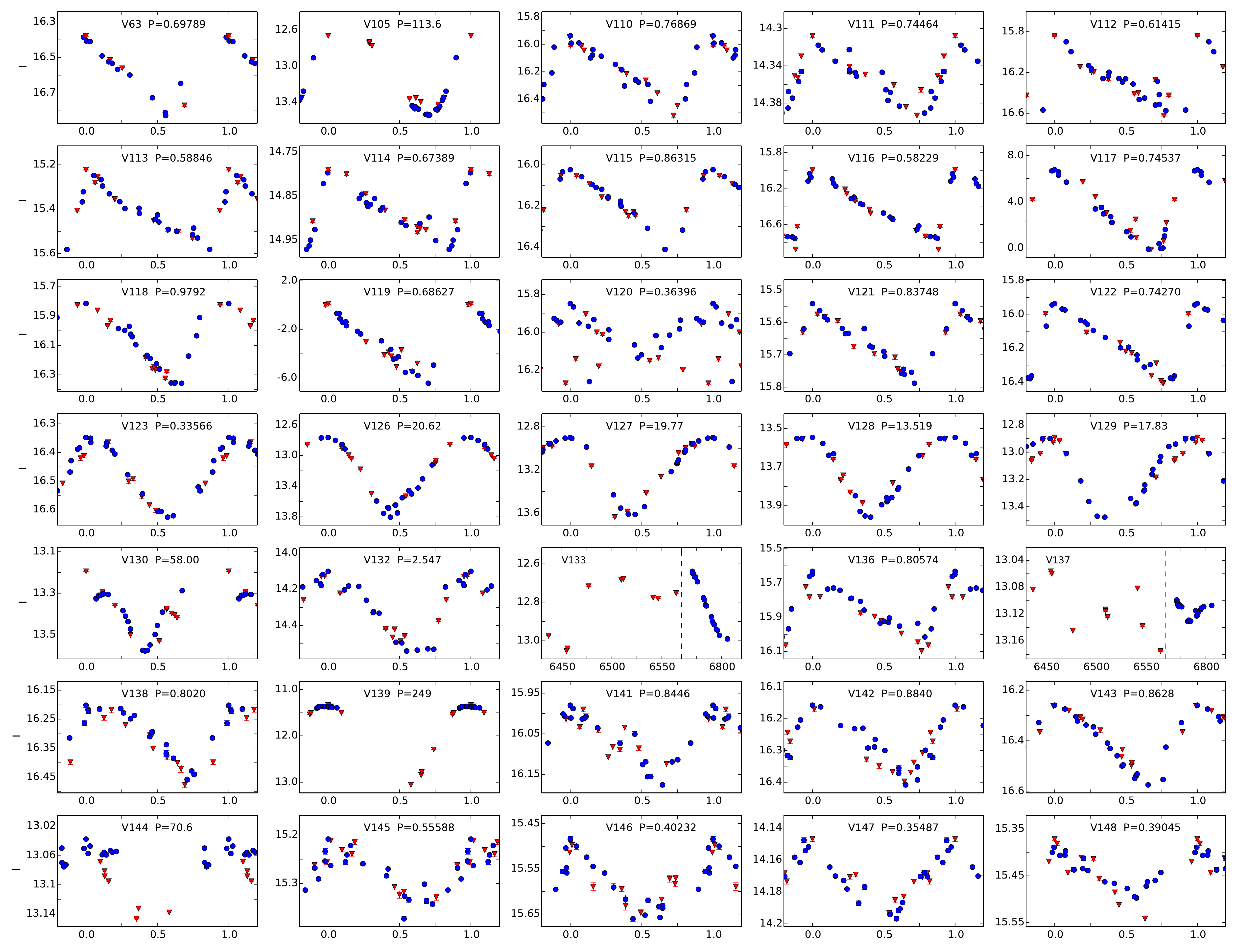}
      \caption{\VarFigText{NGC~6441}{the known} Note that V117 and V119 are plotted in differential flux units, $10^3$ ADU/s, and not calibrated $I$ magnitudes.}
        \label{fig:NGC6441_V}
   \end{figure*}

All of the previously known variables within our FoV are recovered in this analysis.
The light curves for these variables are shown in Fig.~\ref{fig:NGC6441_V}, and their details are listed in Table~\ref{table:NGC6441_V}. The table includes the periods found in P03 and C06.

A discussion of individual variables is given below, but generally it can be noted that for all RRL and CW variables, the same periods are estimated as in P03 and/or C06. For the long period variables there are a few discrepancies. 
A number of the variables are found to be somewhat brighter than in P03. This is most likely due to blending with very close neighbours, leading to an overestimation of the reference flux.

The classification that is given in P03 and/or C06 seems to be correct in almost all cases and have thus been adopted, unless otherwise noted below. A period-luminosity diagram for the stars in the upper part of the instability strip is shown in Fig. \ref{fig:PerLum}, and is discussed in more detail in Sect. \ref{sec:disc6388_6441}.

Satisfactory light curves are found for the three variables that were not detected by C06: V136, V138, and V145. 
Of the five new variables that C06 discovered, three (V146-V148) are within our FoV and have reasonable light curves. 
This highlights the power of EMCCD observations and DIA combined. 

Discussion of individual variables:
\begin{description}
\item[{\bf V63:}]{This RR0 is the only variable that is also found in \cite{Pritzl2001}. }
\item[{\bf V105:}]{We find a slightly shorter period for this SR than in P03.}
\item[{\bf V110, V122:}]{No CMD positions are found for these two variables, but their periods and phased light curves clearly indicate that they are RR0 stars.}
\item[{\bf V111:}]{This RR0 is highly blended with a brighter star, and the mean magnitude is therefore too bright. Due to the strong blending the correct position in the CMD has not been found.}
\item[{\bf V113, V121:}]{Due to blending with nearby stars of similar brightness, these RR0 stars have overestimated mean magnitudes.}
\item[{\bf V114:}]{This RR0 star is heavily blended with a brighter star.}
\item[{\bf V117, V119:}]{Both stars are in a very crowded area and their positions and reference fluxes were not found by the pipeline. 
The correct positions of the stars were found in the summed difference image, and the differential fluxes have been measured for these positions in the difference images. Therefore, as no reference flux could be measured, magnitudes and amplitudes are not given for either of these RR0 stars.}
\item[{\bf V118:}]{Using the period of this star, P03 and C06 can not give a certain classification, but suggest RR0 or CW. 
Based on the position of the star in the CMD $[0.82,16.9]$ we classify it as an RR0 star, although the period is unusually long for this type of star.}
\item[{\bf V120:}]{This variable is classified as an RR1 in both P03 and C06.
Due to the scatter in our light curve and the position of the star in the CMD $[0.60,17.3]$, we have analysed the light curve for any secondary periods, but none have been found. We therefore support the RR1 classification.}
\item[{\bf V126-V129:}]{In P03 these stars are described as CWA candidates. The positions that these stars occupy in the CMD $[(1.0-1.3),(14.0-15.5)]$, seem to support that they are indeed CWA stars.}
\item[{\bf V130:}]{The period of P03 does not phase our light curve well and our best fit period is significantly longer.
Some of the data points seem to have a sort of systematic scatter and it may therefore be an L star, which is supported by its position in the CMD, $[1.64,15.0]$.}
\item[{\bf V132:}]{This star is quite heavily blended, which is reflected in a rather scattered light curve and the fact that we found a mean $I$ magnitude that is about a magnitude higher than what was found in P03. Based on its position in the CMD and the relation between period and luminosity, we classify this star as a CWB.}
\item[{\bf V133, V137:}]{For these two stars it is not possible to phase the light curves properly with the periods given in P03, or any other period, and this suggests that they are L stars. No period is therefore given for these two variables.}
\item[{\bf V136, V138:}]{These two RR0 stars were not found in the C06 analysis. Our data are properly phased by the periods of P03.}
\item[{\bf V139:}]{This SR phases well with the period found by P03.}
\item[{\bf V144:}]{The period from P03 does phase this SR reasonably well, but the phased light curve still looks a bit peculiar, so it might also be an L star.}
\item[{\bf V145:}]{This star is highly blended with a star of similar brightness and was not found in the C06 paper. P03 classifies this as an RR1.
We are able to find a fundamental and first overtone period of $P_0=0.72082$ and $P_1=0.55588$, respectively. 
This gives a first-overtone to fundamental period ratio of $P_1/P_0=0.7712$, which is only slightly higher than the 'canonical' ratio of $\sim0.745$ \citep{Clement2001}. 
We therefore classify this as a double-mode RRL, which also agrees well with the position of the star in the CMD $[0.64,17.3]$.
Studies of other clusters and dwarf galaxies have shown that the period ratio for RR01 stars generally does not exceed 0.748 \citep[e.g.][figure 13]{Clementini2004}, and the period ratio for this star should therefore be confirmed when more data become available for NGC~6441. } 
\item[{\bf V146-V148:}]{These RR1 stars seems to be blended with multiple stars. However, their periods and light curves, although with some scatter, are consistent with their classification.}
\end{description}

\subsubsection{New variables}

\begin{table*}[ht]
\caption{NGC~6441: Details of the 49 new variables found in the cluster. }
\label{table:NGC6441_NV}      
\centering                          
\begin{tabular}{c c c c c c c c c c}        
\hline\hline                 
Var & RA (J2000.0) & Dec. (J2000.0) & Epoch $(d)$ & P $(d)$ & $< I >$ & $A_{i'+z'}$ & Blend & Classification\\    
\hline                        
V151 & 17:50:11.983 & -37:03:11.63 & 6564.5424 & 0.48716 & 16.46 & 0.80 & iii & RR0 \\
V152 & 17:50:12.843 & -37:03:03.48 & 6564.5424 & 0.9432  &  -    &  -   & i   & RR0? \\
V153 & 17:50:12.767 & -37:03:16.58 & 6541.5439 & 9.89    & 13.72 & 0.15 & iii & CWA \\
V154 & 17:50:13.338 & -37:03:01.06 & 6455.7466 & 10.83   & 13.57 & 0.15 & iii & CWA \\
V155 & 17:50:13.231 & -37:03:10.93 & 6564.5424 & 11.45   & 13.03 & 0.04 & iii & ?  \\
V156 & 17:50:11.357 & -37:03:13.57 & 6789.7942 & 11.76   & 13.97 & 0.03 & iii & SR\tablefootmark{a} \\
V157 & 17:50:11.670 & -37:03:14.69 & 6775.8182 & 15.10   & 13.71 & 0.04 & iii & SR\tablefootmark{a} \\
V158 & 17:50:13.637 & -37:02:48.90 & 6784.7970 & 17.50   & 13.89 & 0.05 & iii & SR\tablefootmark{a} \\
V159 & 17:50:13.147 & -37:03:06.55 & 6782.7959 & 18.5    & 13.03 & 0.06 & ii  & SR\tablefootmark{a} \\
V160 & 17:50:13.536 & -37:03:20.37 & 6795.7897 & 20.3    & 13.31 & 0.11 & iii & SR \\
V161 & 17:50:12.645 & -37:03:12.61 & 6564.5424 & 24.6    & 12.95 & 0.09 & iii & SR \\
V162 & 17:50:14.220 & -37:02:50.97 & 6770.8270 & 29.3    & 13.68 & 0.06 & iii & SR \\
V163 & 17:50:13.742 & -37:03:03.30 & 6546.5297 & 41.03   & 12.59 & 0.11 & iii & SR? \\
V164 & 17:50:11.919 & -37:03:11.12 & 6564.5424 & 41.14   & 14.37 & 0.27 & ii  & ?  \\
V165 & 17:50:12.027 & -37:03:10.81 & 6564.5424 & 51.6    & 12.86 & 0.21 & iii & SR \\
V166 & 17:50:13.434 & -37:02:54.72 & 6546.5297 & 52      & 12.83 & 0.17 & iii & SR \\
V167 & 17:50:14.037 & -37:03:02.88 & 6789.7942 & 86      & 12.78 & 0.13 & iii & SR \\
V168 & 17:50:13.980 & -37:03:12.93 & 6789.7942 & 128     & 13.13 & 0.11 & iii & SR \\
V169 & 17:50:13.918 & -37:03:25.06 &  -  &  -  & 12.99 & 0.68 & iii & L  \\
V170 & 17:50:13.625 & -37:03:08.13 &  -  &  -  & 12.89 & 0.48 & iii & L  \\
V171 & 17:50:13.065 & -37:02:58.75 &  -  &  -  & 12.71 & 0.29 & iii & L  \\
V172 & 17:50:14.561 & -37:03:14.81 &  -  &  -  & 13.06 & 0.21 & iii & L  \\
V173 & 17:50:14.253 & -37:03:04.74 &  -  &  -  & 13.10 & 0.16 & iii & L  \\
V174 & 17:50:12.059 & -37:02:51.54 &  -  &  -  & 13.06 & 0.14 & iii & L  \\
V175 & 17:50:11.371 & -37:03:23.18 &  -  &  -  & 13.27 & 0.14 & iii & L  \\
V176 & 17:50:12.222 & -37:03:04.85 &  -  &  -  & 12.65 & 0.11 & iii & L  \\
V177 & 17:50:14.376 & -37:03:19.09 &  -  &  -  & 13.35 & 0.09 & iii & L  \\
V178 & 17:50:12.930 & -37:02:56.26 &  -  &  -  & 12.91 & 0.08 & iii & L  \\
V179 & 17:50:12.822 & -37:03:04.36 &  -  &  -  & 12.72 & 0.07 & ii  & L  \\
V180 & 17:50:12.433 & -37:03:12.24 &  -  &  -  & 13.72 & 0.07 & iii & L  \\
V181 & 17:50:14.294 & -37:03:15.36 &  -  &  -  & 13.73 & 0.07 & iii & L  \\
V182 & 17:50:13.165 & -37:03:08.78 &  -  &  -  & 12.62 & 0.06 & iii & L  \\
V183 & 17:50:11.939 & -37:02:52.42 &  -  &  -  & 12.72 & 0.06 & iii & L  \\
V184 & 17:50:12.906 & -37:03:19.30 &  -  &  -  & 12.84 & 0.06 & iii & L  \\
V185 & 17:50:11.845 & -37:03:00.62 &  -  &  -  & 13.07 & 0.06 & iii & L  \\
V186 & 17:50:13.911 & -37:03:15.79 &  -  &  -  & 13.17 & 0.06 & iii & L  \\
V187 & 17:50:12.583 & -37:02:49.07 &  -  &  -  & 13.25 & 0.06 & iii & L  \\
V188 & 17:50:12.364 & -37:03:24.91 &  -  &  -  & 13.70 & 0.06 & iii & L  \\
V189 & 17:50:14.453 & -37:03:08.15 &  -  &  -  & 13.81 & 0.06 & iii & L  \\
V190 & 17:50:13.032 & -37:03:00.01 &  -  &  -  & 13.04 & 0.05 & iii & L  \\
V191 & 17:50:13.309 & -37:03:04.07 &  -  &  -  & 12.46 & 0.04 & iii & L  \\
V192 & 17:50:13.406 & -37:03:06.76 &  -  &  -  & 12.80 & 0.04 & iii & L  \\
V193 & 17:50:12.269 & -37:03:10.06 &  -  &  -  & 13.19 & 0.04 & iii & L  \\
V194 & 17:50:13.902 & -37:03:05.68 &  -  &  -  & 13.19 & 0.04 & ii  & L  \\
V195 & 17:50:13.649 & -37:02:53.87 &  -  &  -  & 13.41 & 0.04 & iii & L  \\
V196 & 17:50:11.542 & -37:02:56.76 &  -  &  -  & 13.81 & 0.04 & iii & L  \\
V197 & 17:50:11.861 & -37:02:49.20 &  -  &  -  & 14.07 & 0.04 & iii & L  \\
V198 & 17:50:13.173 & -37:02:56.04 &  -  &  -  & 13.13 & 0.03 & iii & L  \\
V199 & 17:50:12.346 & -37:02:57.29 &  -  &  -  & 13.46 & 0.03 & iii & L  \\
\hline                                   
\end{tabular}
\tablefoot{\varTableText{2456541.54}{}
\tablefoottext{a}{These stars have shorter periods than expected for stars of the SR class, but it is not clear how to otherwise classify them according to the GCVS schema given their position on the RGB.}}
\end{table*}

\begin{figure*}[ht]
   \centering
   \includegraphics[width=\linewidth]{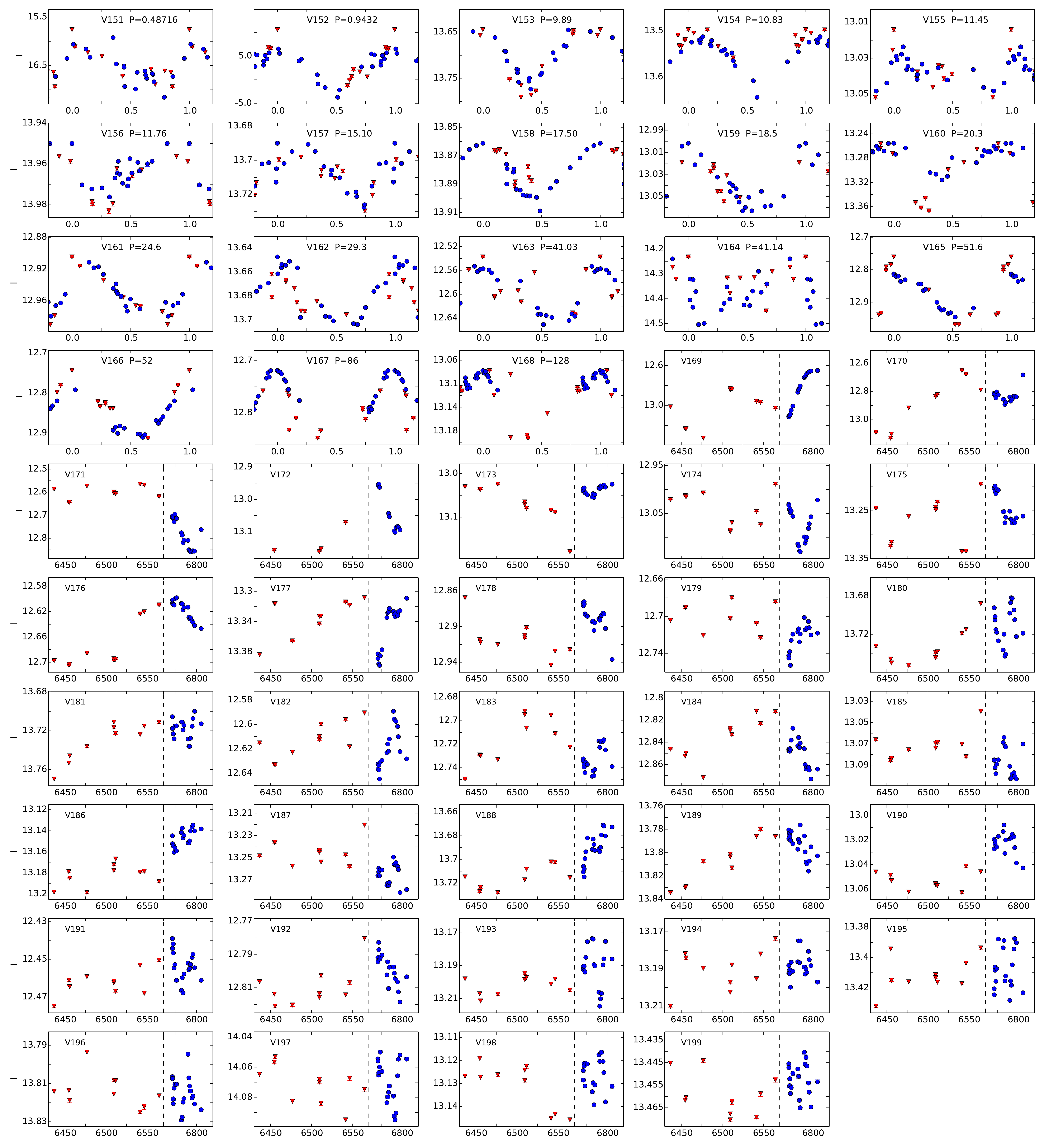}
      \caption{\VarFigText{NGC~6441}{the new} Note that V152 is plotted in differential flux units, $10^3$ ADU/s, and not calibrated $I$ magnitudes.}
        \label{fig:NGC6441_NV}
   \end{figure*}

In this study we were able to find 49 new variable stars for NGC~6441. These are all listed in Table~\ref{table:NGC6441_NV} and their light curves are shown in Fig.~\ref{fig:NGC6441_NV}.

Similar to NGC~6388, most of the new variables are RGB stars, many with small amplitudes, which is probably the reason why they have not been detected until now. 
There are, however, also one (possibly two) previously undetected RRL stars and two new CW stars. 
For many of the new long period variables it has been hard to determine a period, which could indicate that they are irregular. 

It should be noted that three of the new variables (V151, V164, and V165) are located very close to each other. Resolving these three variables would have been difficult using conventional imaging, as they are all within a radius of $\sim 0\farcs9$.

Discussion of individual variables:
\begin{description}
\item[{\bf V151:}]{From the period and position in the CMD $[0.6,17.0]$ this star can be safely classified as an RRL star, and the asymmetry of the light curve suggests that it is most likely an RR0. 
The amplitude listed is very large compared to other RRL stars, and there seems to be some scatter in the light curve (caused by its proximity to three other brighter variables) so the actual amplitude is probably about $A_{i'+z'} \sim 0.8$ mag.}
\item[{\bf V152:}]{This star is highly blended with a star of similar brightness and lies very close to another variable star, V179. The position and reference flux were therefore not found by the pipeline, as in the cases of V117 and V119. The positions and differential fluxes were found manually, but again the reference flux is not measured, so no magnitude or amplitude is given. Based on the period, light curve shape, and CMD position $[0.85,17.4]$, we classify this tentatively as an RR0 star.}
\item[{\bf V153, V154:}]{Based on their periods, light curves, and positions in the CMD $[1.05,15.0]$, we classify these two stars as CWA stars.}
\item[{\bf V155:}]{Both the period and magnitude suggest that this is either a CW or an SR star, but the position of this star in the CMD is $[-0.5,13.5]$, which means that it is far too blue to fit any of these classifications. We do not attempt to classify this variable. }
\item[{\bf V156-V162, V165-V168:}]{The CMD puts all of these stars on the RGB. 
As it has been possible to estimate periods for them, we classify these as SR stars.
}
\item[{\bf V163:}]{This star has no CMD information, but based on the period, light curve, and magnitude, it is most likely an SR star.}
\item[{\bf V164:}]{The CMD position $[1.1,15.5]$ and the mean magnitude of this star would normally imply that it is a CW star. However the light curve is scattered and the derived period uncertain. We refrain from classifying this variable. }
\item[{\bf V169-V199:}]{We have classified these RGB stars for which we have been unable to derive periods as L stars.}
\end{description}

\subsection{NGC~6528}
\subsubsection{Background information}
According to \citet{SawyerHogg1973} there are a few variables from the rich Galactic field projected against the cluster, but none are considered to be cluster members.
As NGC~6528 might be the most metal-rich cluster in the Galaxy, it has been studied quite extensively in a number of photometric studies \citep[e.g.][]{Ortolani1992,Richtler1998,Feltzing2002,Calamida2014}. 
However, so far no variable stars have been reported.
\citet{Ortolani1992} mention that the large spread in the RGB that is found may be due to variability, but this is not mentioned in any later article.

\subsubsection{New Variables}

\begin{figure}[t]
   \centering
   \includegraphics[width=\linewidth]{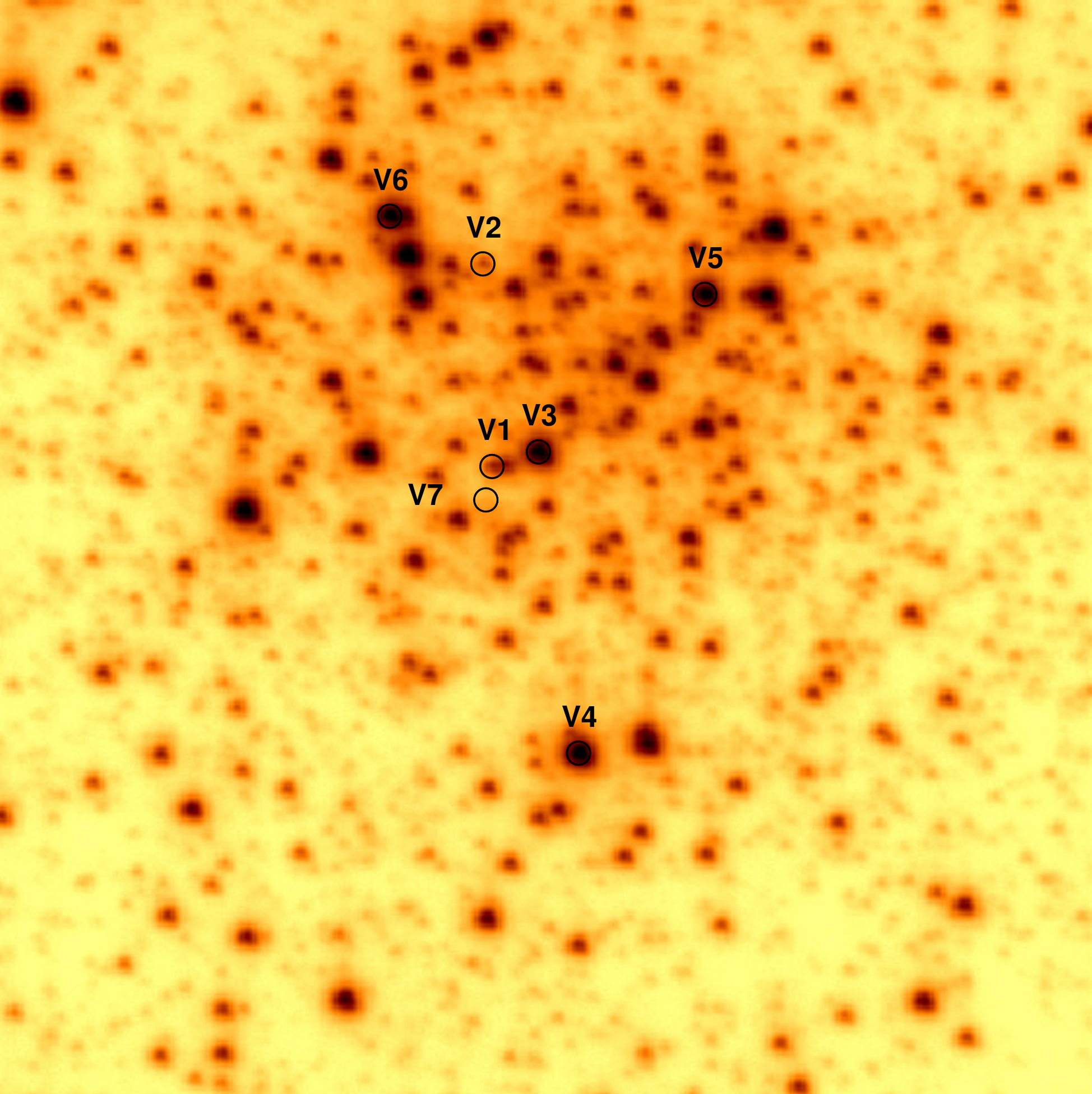}
      \caption{\fchartText{NGC~6528}{}{41}{41}}
        \label{fig:NGC6528_ref}
   \end{figure}

\begin{figure}[t]
   \centering
   \includegraphics[width=\linewidth]{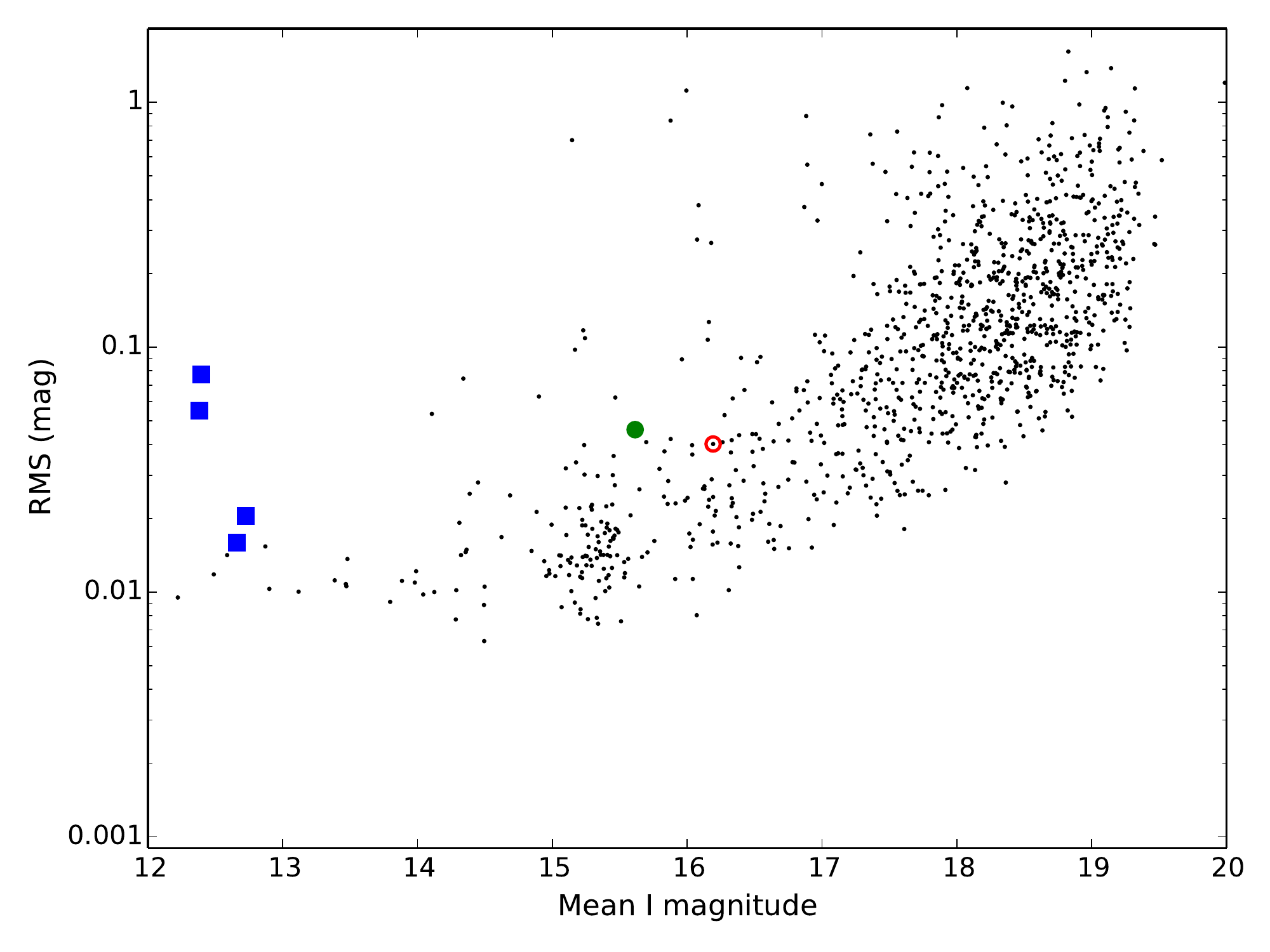}
      \caption{NGC~6528: Plot of the RMS magnitude deviation versus the mean magnitude for each of the 1103 calibrated $I$ light curves. The stars that show variability in our study are plotted with the same symbols as in Fig. \ref{fig:NGC6388_cmd}.}
        \label{fig:NGC6528_rms}
   \end{figure}
   
A finding chart for the cluster containing the variables detected in this study is shown in Fig.~\ref{fig:NGC6528_ref}, and Fig.~\ref{fig:NGC6528_rms} shows the RMS magnitude deviation for the 1103 stars with calibrated $I$ magnitudes versus their mean magnitude. 

\begin{table*}
\caption{NGC~6528: Details of the 7 new variables found in the cluster. }
\label{table:NGC6528_NV}      
\centering                          
\begin{tabular}{c c c c c c c c c}        
\hline\hline                 
Var & RA (J2000.0) & Dec. (J2000.0) & Epoch $(d)$ & P $(d)$ & $< I >$ & $A_{i'+z'}$ & Blend & Classification\\    
\hline                        
V1 & 18:04:49.380 & -30:03:27.25 & 6477.8478 & 0.25591 & 15.61 & 0.19 & i & RR1  \\
V2 & 18:04:49.368 & -30:03:19.53 & 6541.5615 & 0.8165 & 16.18 & 0.14 & iii & RR0?  \\
V3 & 18:04:49.518 & -30:03:26.74 &  -  &  -  & 12.46 & 0.25 & iii & L  \\
V4 & 18:04:49.614 & -30:03:38.25 &  -  &  -  & 12.40 & 0.21 & iii & L  \\
V5 & 18:04:50.016 & -30:03:20.90 &  -  &  -  & 12.70 & 0.10 & iii & L  \\
V6 & 18:04:49.099 & -30:03:17.64 &  -  &  -  & 12.66 & 0.06 & iii & L  \\
V7 & 18:04:49.360 & -30:03:28.52 &  -  &  -  & 19.6  &  -   & -   & E?  \\
\hline                                   
\end{tabular}
\tablefoot{\varTableText{2456510.63}{}}
\end{table*}

\begin{figure}[ht]
   \centering
   \includegraphics[width=\linewidth]{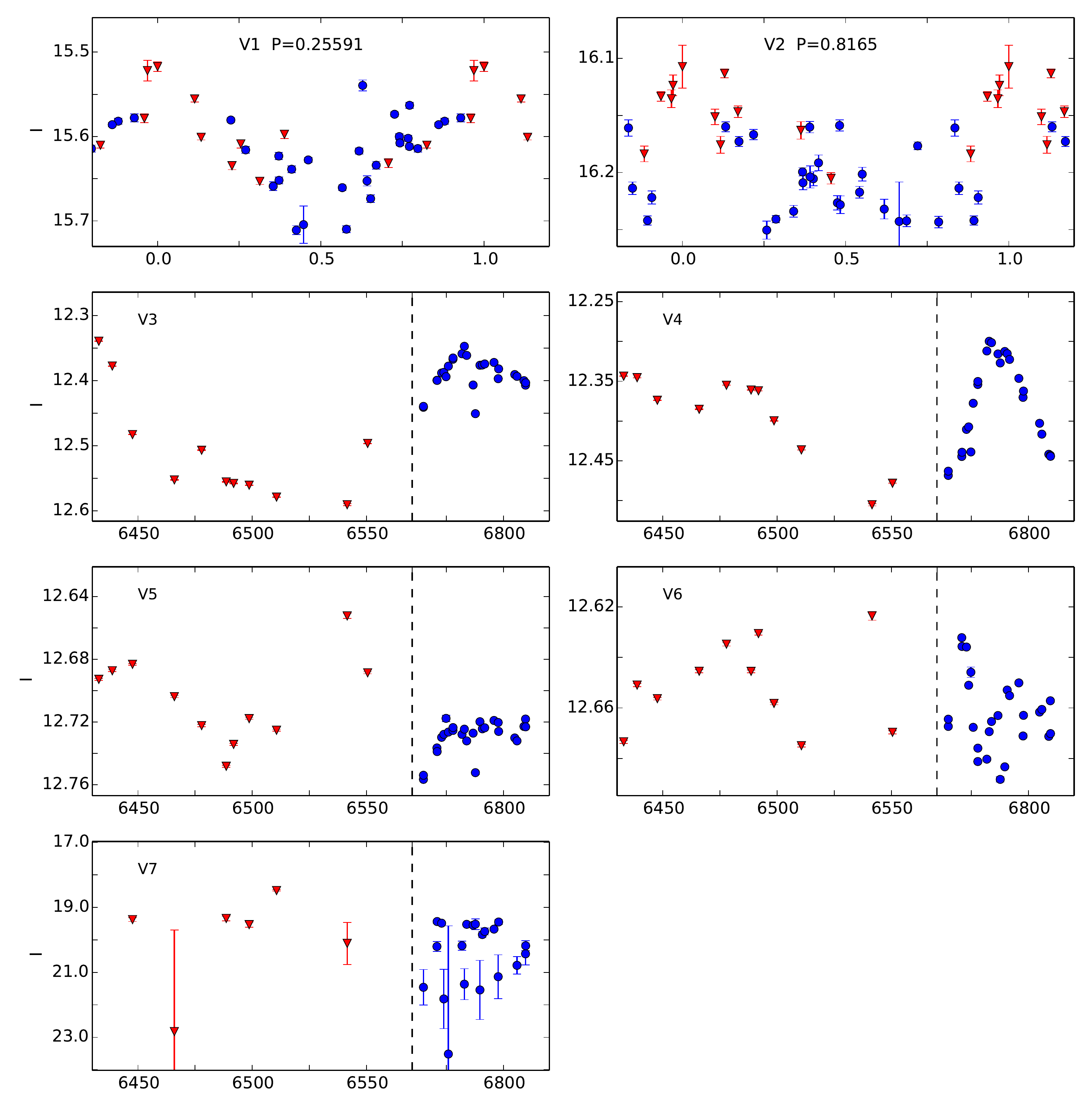}
      \caption{\VarFigText{NGC~6528}{the}}
        \label{fig:NGC6528_NV}
   \end{figure}

We are able to find seven new variable stars for NGC~6528. 
Unfortunately there is no CMD for this cluster, which complicates the classification of the variables, but we classify one (possibly two) as RRL stars and four as long period irregular stars.
The light curves for these variables are shown in Fig.~\ref{fig:NGC6528_NV} and their details are listed in Table~\ref{table:NGC6528_NV}. 

\begin{description}
\item[{\bf V1:}]{The period of this star suggests that it is an RR1, and the magnitude indicates that it is highly blended with a brighter star. 
Due to the scatter in the light curve, we have analysed it for secondary periods, but none were found.}
\item[{\bf V2:}]{The period of this star indicates that it is a RR0, even though the light curve looks somewhat noisy. 
Since the amplitude is somewhat smaller than what we expect for an RR0 star, we leave our classification as tentative.}
\item[{\bf V3-V6:}]{These four stars are most likely L stars, based on their magnitudes and that it has not been possible to find any periods that phase their light curves in a reasonable way.}
\item[{\bf V7:}]{The mean magnitude of this star is at the limit of our detection threshold, and it disappears in $\sim30\%$ of our images. It is probably an eclipsing binary (E), but we refrain from making a firm classification.}
\end{description}

\subsection{NGC~6638}
The first 19 variable stars in this cluster were found by \citet{Terzan1968}. 
Of these, four were classified as Mira variables with periods between 156 and 279 days, and the remaining ones were neither classified nor had their periods estimated.
\citet{SawyerHogg1974} were able to discover the variability of a further 26 stars from a photographic collection made in 1939 and 1972. 
All of these 45 variable stars are distributed in a wide field of $30^{\prime} \times 30^{\prime}$ and only a few are located close to the central parts of the cluster. 

In \citet{Rutily1977} (hereafter R77) a total of 63 variables are presented. That article contains finding charts and periods for many of the variables. 
As NGC~6638 does not have a very dense central region, it was possible to detect variable stars quite close to the centre, but no periods for the central stars are given.

\begin{figure}[t]
   \centering
   \includegraphics[width=\linewidth]{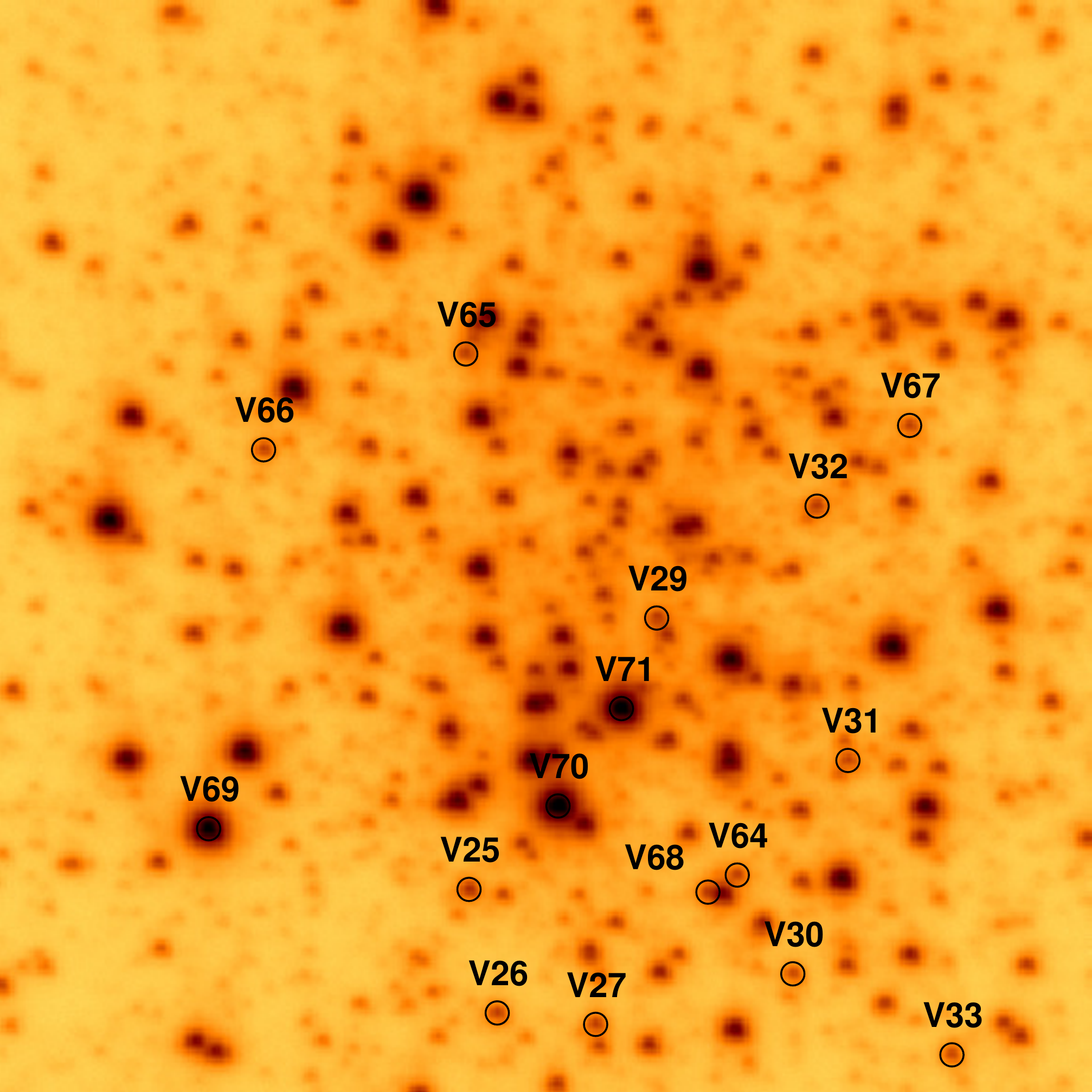}
      \caption{\fchartText{NGC~6638}{}{41}{41}}
        \label{fig:NGC6638_ref}
   \end{figure}

\begin{figure}[t]
   \centering
   \includegraphics[width=\linewidth]{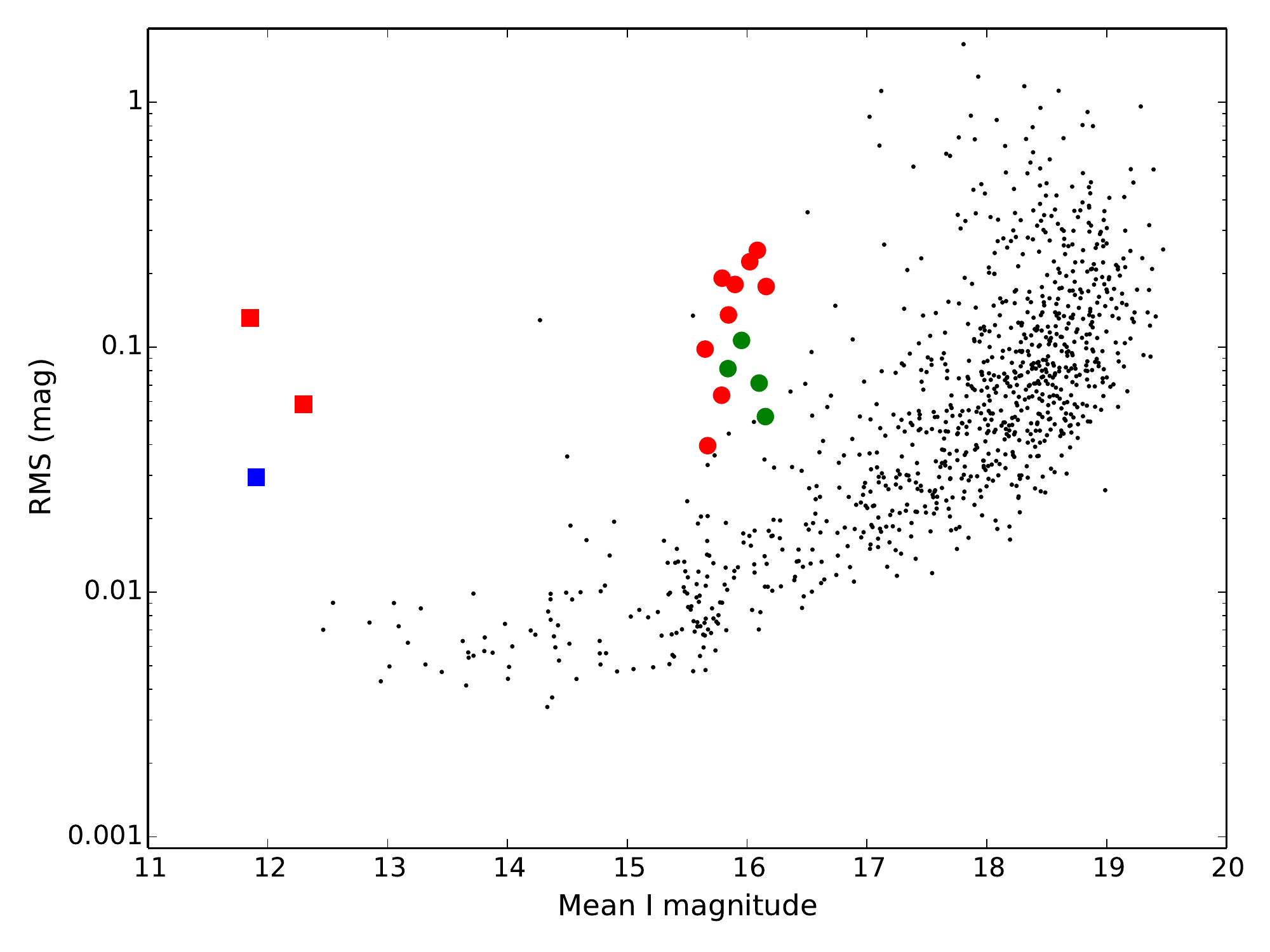}
      \caption{NGC~6638: Plot of the RMS magnitude deviation versus the mean magnitude for each of the 981 calibrated I light curves. The stars that show variability in our study are plotted with the same symbols as in Fig. \ref{fig:NGC6388_cmd}.} 
        \label{fig:NGC6638_rms}
   \end{figure}
   
A finding chart for the cluster containing the variables detected in this study is shown in Fig.~\ref{fig:NGC6638_ref}. 
Fig.~\ref{fig:NGC6638_rms} shows the RMS magnitude deviation for the 981 stars with calibrated $I$ magnitude versus their mean magnitude.

\subsubsection{Known variables}
\begin{table*}
\caption{NGC~6638: Details of the 8 previously known variables in our FoV. }
\label{table:NGC6638_V}      
\centering                          
\begin{tabular}{c c c c c c c c c}        
\hline\hline                 
Var & RA (J2000.0) & Dec. (J2000.0) & Epoch $(d)$ & P $(d)$  & $< I >$ & $A_{i'+z'}$ & Blend & Classification\\ 
\hline                        
V25 & 18:30:55.564 & -25:30:04.05 & 6511.6617 & 0.67276 & 15.79 & 0.66 &  iii & RR0 \\
V26 & 18:30:55.635 & -25:30:08.79 & 6782.8404 & 0.66743 & 15.85 & 0.47 &  iii & RR0 \\
V27 & 18:30:55.912 & -25:30:09.31 & 6773.8921 & 0.59969 & 15.87 & 0.68 &  iii & RR0 \\
V29 & 18:30:56.111 & -25:29:53.87 & 6773.8921 & 0.257893 & 15.92 & 0.31 & iii & RR1  \\
V30 & 18:30:56.471 & -25:30:07.56 & 6458.6160 & 0.50650 & 16.00 & 0.75 &  iii & RR0 \\
V31 & 18:30:56.640 & -25:29:59.47 & 6783.8751 & 0.45795 & 15.88 & 0.67 &  iii & RR0 \\
V32 & 18:30:56.570 & -25:29:49.75 & 6773.8921 & 0.56830 & 15.65 & 0.35 &  iii & RR0  \\
V33 & 18:30:56.914 & -25:30:10.79 & 6791.7951 & 0.32378 & 16.07 & 0.22 &  iii & RR1  \\
\hline                                   
\end{tabular}
\tablefoot{\varTableText{2456511.65}{}}
\end{table*}

\begin{figure*}[t]
   \centering
   \subfloat[][Known variables]{\centering
                \includegraphics[width=\linewidth]{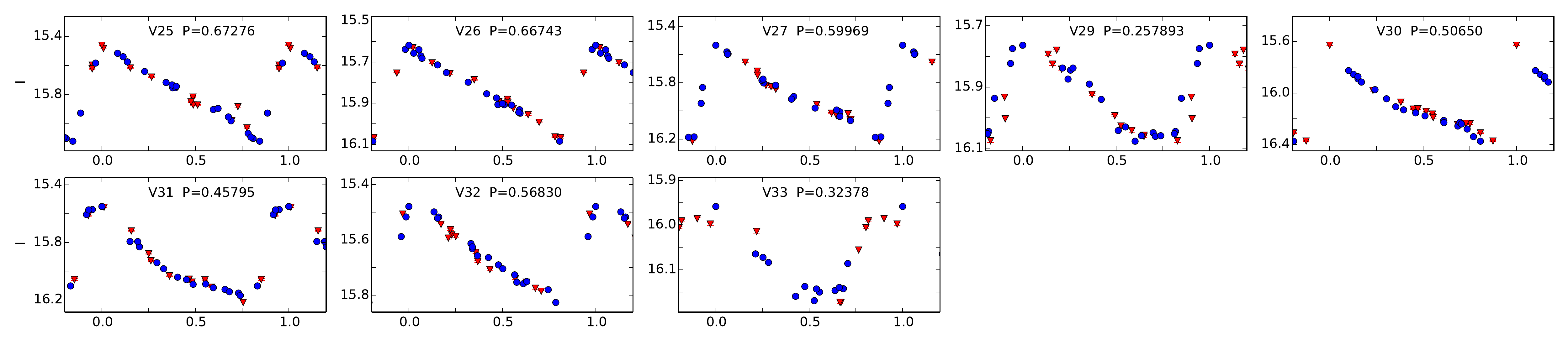}
                \label{fig:NGC6638_V}} \newline
   \vspace{-0.5pt}
   \subfloat[][New variables]{\centering
                \includegraphics[width=\linewidth]{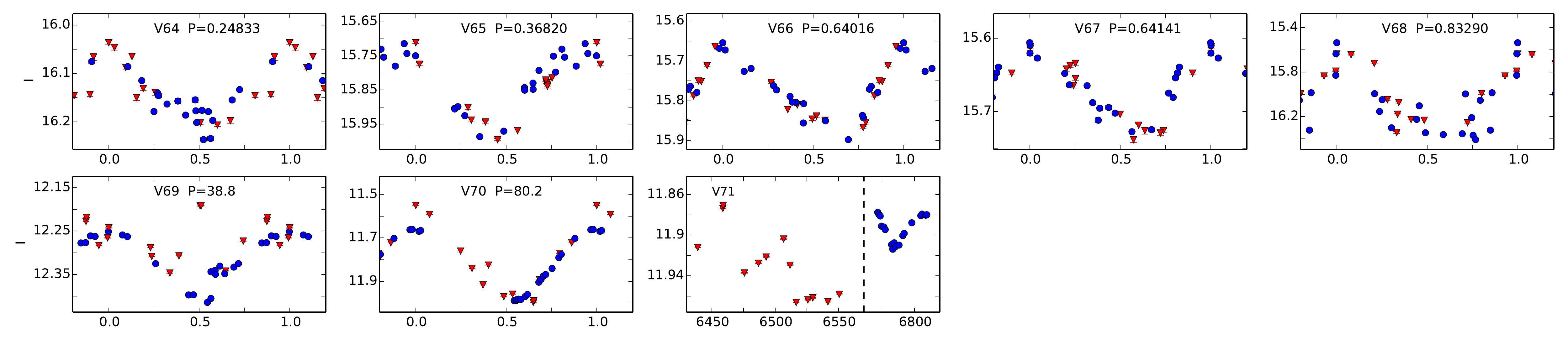}
                \label{fig:NGC6638_NV}}
   
      \caption{NGC~6638: Light curves for the variables in our FoV. Red triangles are 2013 data and blue circles are 2014 data. Error bars are plotted but are smaller than the data symbols in many cases. Light curves with confirmed periods are phased. For those variables without periods the x-axis are in (HJD - 2450000), and the dashed line indicates that the period from HJD 2456600 to 2456750 has been removed from the plot, as no observations were performed during this time range. }
        \label{fig:NGC6638_Var}
   \end{figure*}

All of the previously known variable stars within our FoV have been located and confirmed as variables, and this is the first study to present periods and classifications for these stars.
The light curves are plotted in Fig.~\ref{fig:NGC6638_V} and their details can be found in Table~\ref{table:NGC6638_V}.

The fact that no CMD is available for this study makes it a little harder to classify the variables, but with the periods and magnitudes found, we are quite certain that they are all RRL stars. 

\begin{description}
\item[{\bf V25-V27, V30-V32}:]{These stars all have periods, light curve shapes, and amplitudes typical of RR0 stars.}
\item[{\bf V29}:]{The star given as V29 in R77 is not found to be variable. A star located $3\arcsec$ away \emph{is} found to be variable and we believe that this is the actual V29. Based on the period and magnitude, this star is classified as an RR1, despite the unusual asymmetric shape of the light curve.}
\item[{\bf V33}:]{Based on the period, light curve shape, and magnitude, we classify this star as an RR1.}
\end{description}

\subsubsection{New variables}

\begin{table*}
\caption{NGC~6638: Details of the 8 new variables found in the cluster.}
\label{table:NGC6638_NV}      
\centering                          
\begin{tabular}{c c c c c c c c c c}        
\hline\hline                 
Var & RA (J2000.0) & Dec. (J2000.0) & Epoch $(d)$ & P $(d)$ & $< I >$ & $A_{i'+z'}$ & Blend & Classification\\    
\hline                        
V64 & 18:30:56.321 & -25:30:03.74 & 6541.5840 & 0.24833 & 16.14 & 0.20 &  iii & RR1  \\
V65 & 18:30:55.591 & -25:29:43.63 & 6458.8425 & 0.36820 & 15.85 & 0.28 &  iii & RR1  \\
V66 & 18:30:55.015 & -25:29:47.10 & 6783.8751 & 0.64016 & 15.78 & 0.24 &  iii & RR0  \\
V67 & 18:30:56.836 & -25:29:46.75 & 6783.8751 & 0.64141 & 15.67 & 0.13 &  iii & RR0  \\
V68 & 18:30:56.237 & -25:30:04.37 & 6797.8372 & 0.83290 & 15.97 & 0.87 &  i   & RR0  \\
V69 & 18:30:54.834 & -25:30:01.50 & 6516.5499 & 38.8 & 12.30 & 0.22    &  iii & SR  \\
V70 & 18:30:55.820 & -25:30:00.95 & 6486.7350 & 80.2 & 11.77 & 0.45    &  iii & SR  \\
V71 & 18:30:56.005 & -25:29:57.29 &  -  &  -  & 11.92 & 0.10           &  iii & L  \\
\hline                                   
\end{tabular}
\tablefoot{\varTableText{2456511.65}{}}
\end{table*}


In this study we were able to find eight new variable stars for NGC~6638. The lack of a CMD complicates the classification, but we classify five as RRL stars and three as long period variables. 
The light curves for these variables are shown in Fig.~\ref{fig:NGC6638_NV} and their details are listed in Table~\ref{table:NGC6638_NV}. 

Especially interesting are the two variables V64 and V68, that are located very close to the same bright star (within $1\farcs2$ and $0\farcs8$, respectively). Using conventional imaging it would have been hard to distinguish these three stars from each other, but with the high resolution we achieve with the EMCCD camera this is possible.

\begin{description}
\item[{\bf V64, V65:}]{Based on the period, light curve shape, and magnitude, these stars can be safely classified as RR1. }
\item[{\bf V66-V68:}]{Three stars with periods and light curve shapes, that strongly indicate that they are RR0 stars.}
\item[{\bf V69:}]{The period found for this SR phases the light curve reasonably well, but there are some outlier points, which might indicate that it is actually irregular.}
\item[{\bf V70:}]{The light curve of this SR is phased well by the period.}
\item[{\bf V71:}]{No period was found for this star, and it is therefore classified as an L star.}
\end{description}

\subsection{NGC~6652}
\subsubsection{Background information}
Using 23 photographic plates taken with the 1m Yale telescope at CTIO in 1977, \citet{Hazen1989} found 24 variable stars in this cluster. Nine variables (V1-9) were found within the tidal radius of the cluster, and 15 outside it. 
Two of the variables (V7 and V9) had already been published by \citet{Plaut1971} as part of the Palomar-Groningen variable-star survey.
None of these variable stars are close to the rather dense central part of the cluster nor do they lie within our FoV. 

There are 12 known X-ray sources in NGC~6652 \citep{Deutsch1998,Deutsch2000,Heinke2001,Coomber2011,Stacey2012,Engel2012}. Sources A-C have been assigned variable numbers V10-V12, respectively, by C. Clement (private communication). 
Seven of the X-ray sources lie within our FoV (sources B-E and G-H). Only sources B and H show signs of variability in our difference images and for the remaining sources we cannot find any clear optical counterparts in our reference image at the positions given by \citet{Coomber2011} and \citet{Stacey2012}, although we note that our limiting magnitude is $\sim$19.5~mag (see Fig.~\ref{fig:NGC6652_rms}).

\begin{figure}[t]
   \centering
   \includegraphics[width=\linewidth]{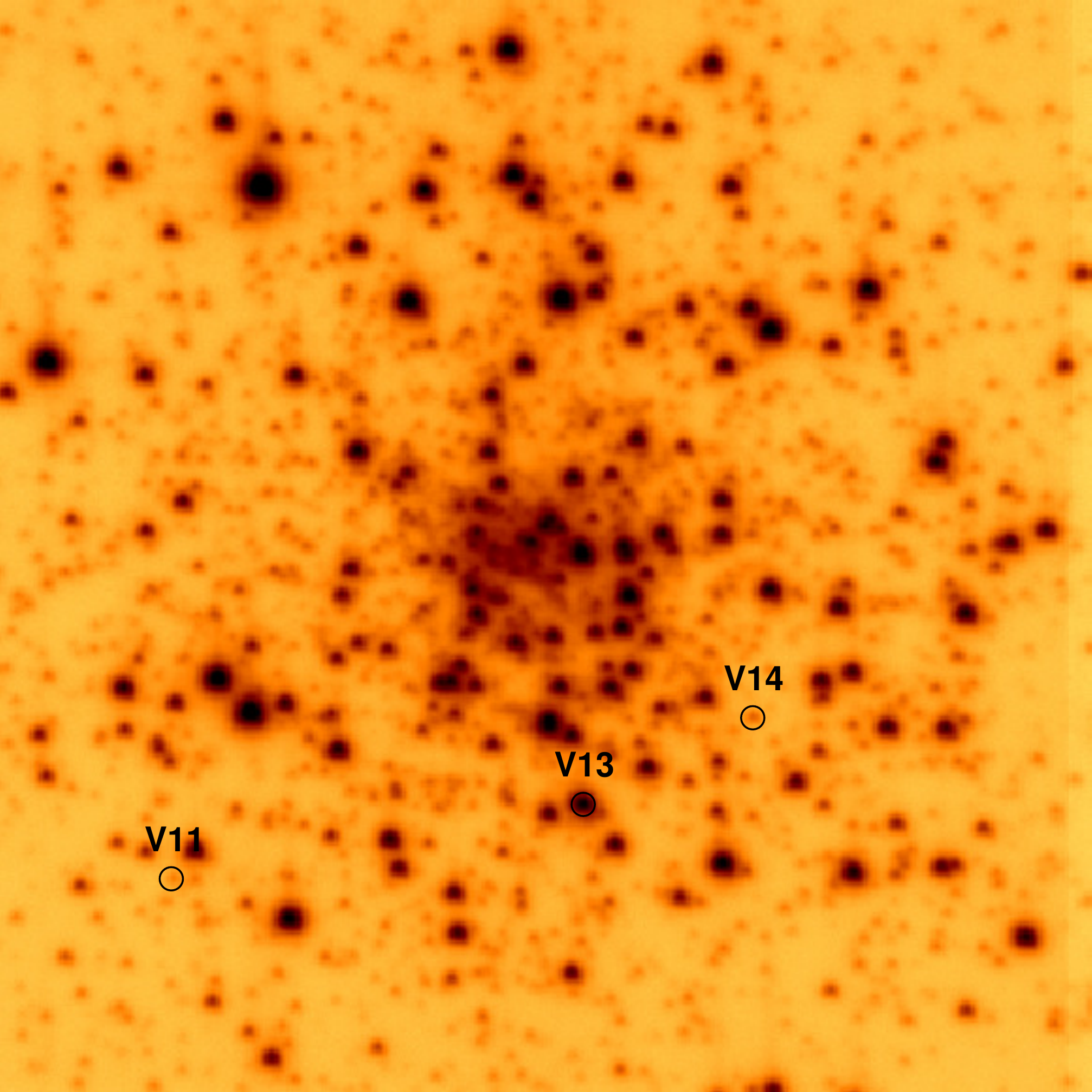}
      \caption{\fchartText{NGC~6652}{}{41}{41}}
        \label{fig:NGC6652_ref}
   \end{figure}

\begin{figure}[t]
   \centering
   \includegraphics[width=\linewidth]{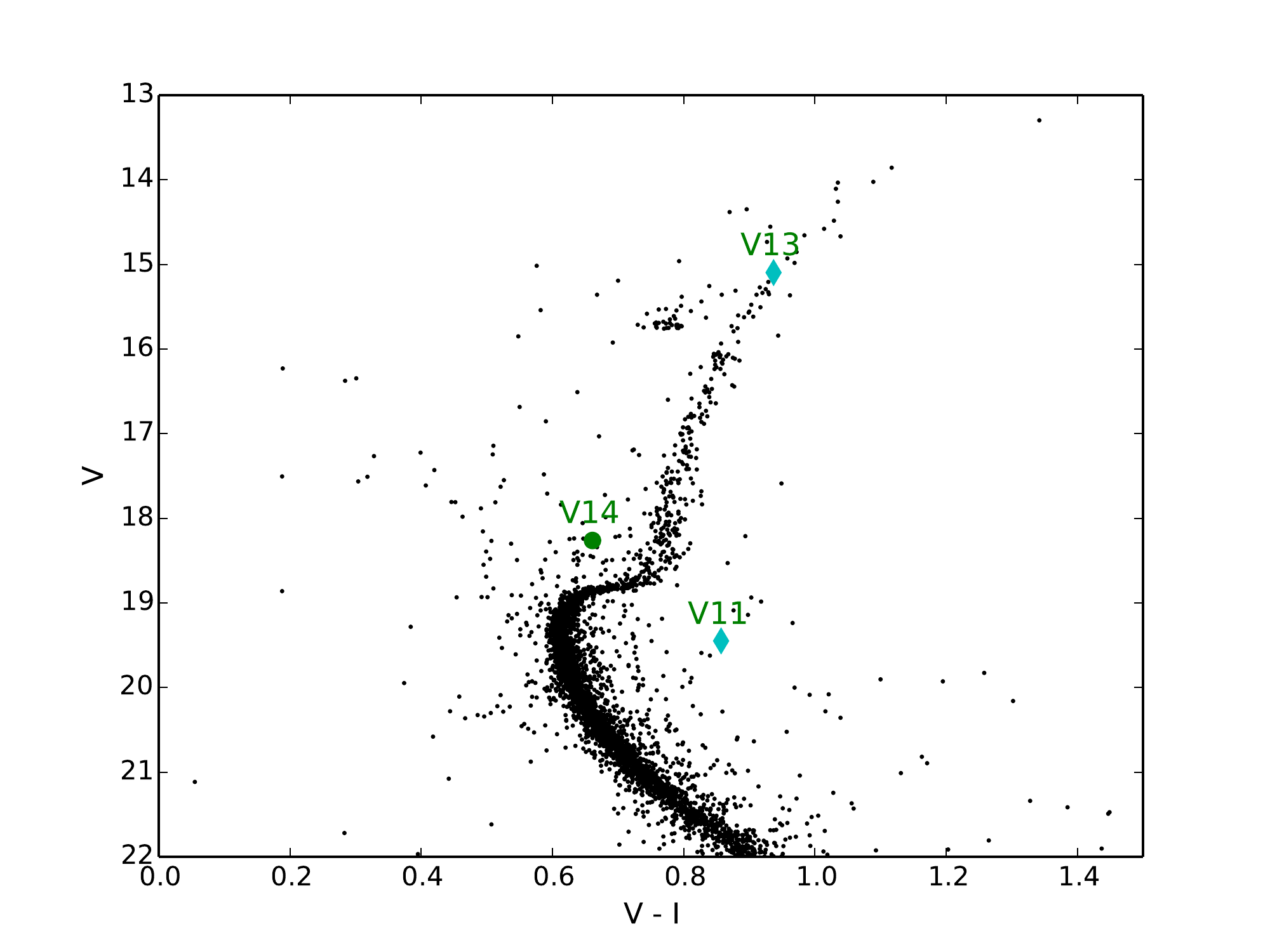}
      \caption{NGC~6652: $(V-I),V$ colour-magnitude diagram made from HST/ACS data as explained in Sect. \ref{sec:cmd}. The three stars that show variability in our study are plotted. The two previously known X-ray sources are marked as cyan diamonds and the new found RR1 is marked with a green circle.}
        \label{fig:NGC6652_cmd}
   \end{figure}

\begin{figure}[t]
   \centering
   \includegraphics[width=\linewidth]{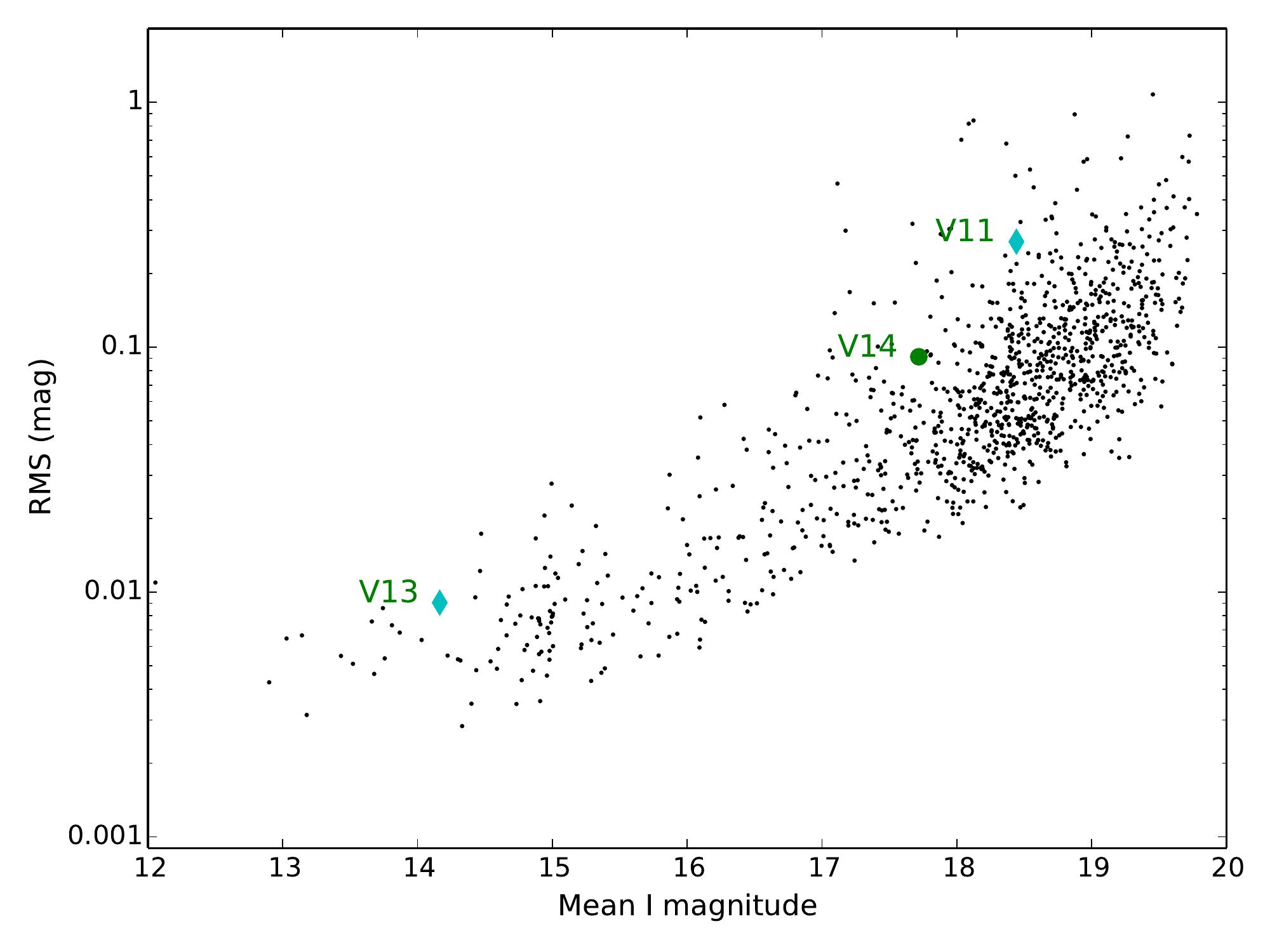}
      \caption{NGC~6652: Plot of the RMS magnitude deviation versus the mean magnitude for each of the 1098 calibrated I light curves. The variables are plotted with the same symbols as in Fig.~\ref{fig:NGC6652_cmd}.
              }
        \label{fig:NGC6652_rms}
   \end{figure}

A finding chart for the cluster containing the variables detected in this study is shown in Fig.~\ref{fig:NGC6652_ref} and Fig.~\ref{fig:NGC6652_cmd} shows a CMD of the stars within our FoV with the variables overplotted.
Fig.~\ref{fig:NGC6652_rms} shows the RMS magnitude deviation for the 1098 stars with calibrated $I$ magnitudes versus their mean magnitude.
The light curves of the variables are plotted in Fig.~\ref{fig:NGC6652_V} and their details may be found in Table~\ref{table:NGC6652_V}. 

\subsubsection{Known variables}
\begin{figure}[t]
   \centering
   \includegraphics[width=\linewidth]{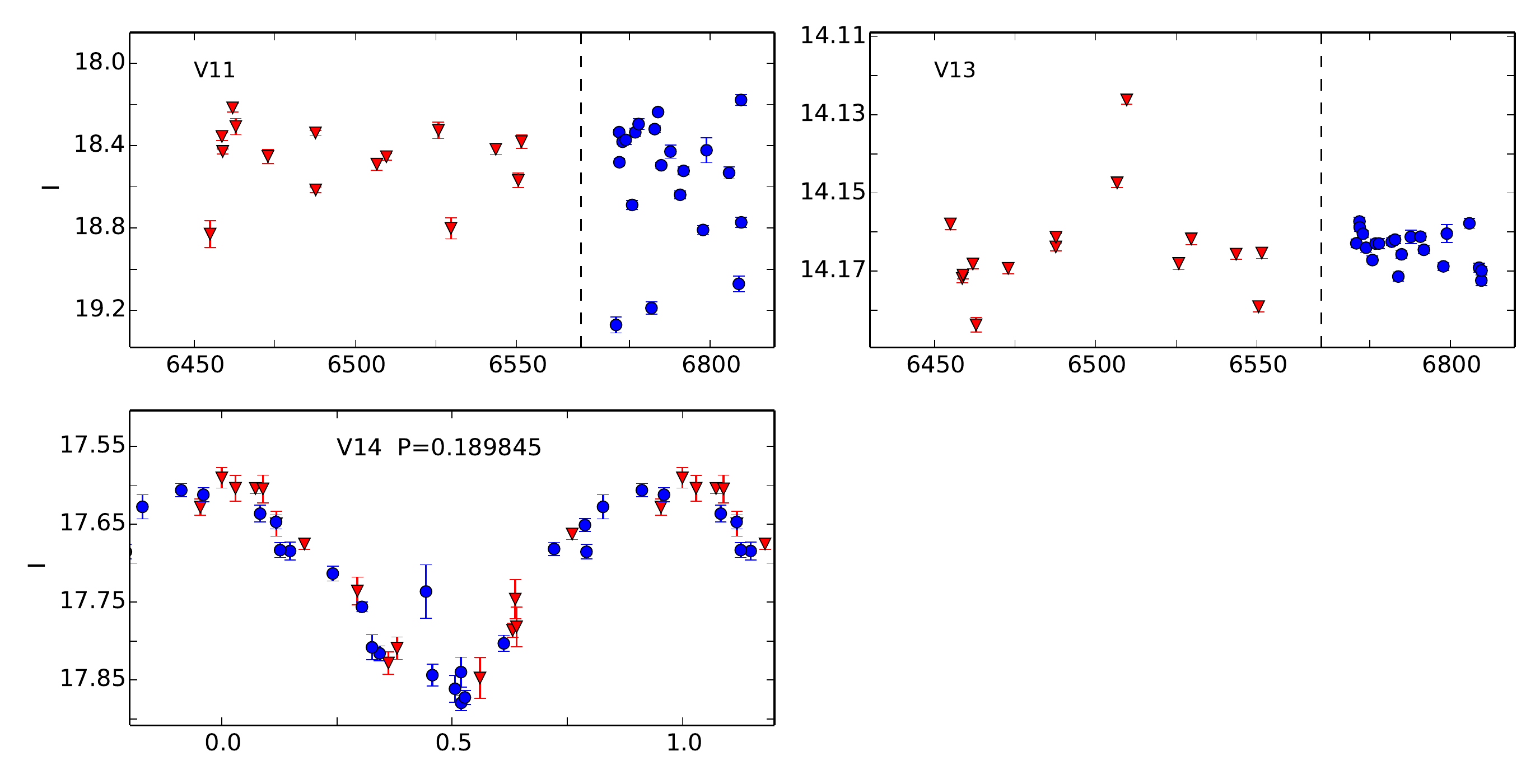}
      \caption{\VarFigText{NGC~6652}{the}}
        \label{fig:NGC6652_V}
   \end{figure}

{\bf V11} (source B) was found by \citet{Coomber2011} to be undergoing rapid X-ray flaring on time scales of less than 100 seconds. Like Engel et al. (2012), we have detected clear variability in the optical counterpart with an amplitude of up to $\sim$1.1~mag (see Fig.~\ref{fig:NGC6652_V}). However, as in previous studies, we cannot find a period on which the light curve can be phased. \citet{Stacey2012} suggest that V11 might be a special type of low mass X-ray binary (LMXB) or a very faint X-ray transient (VFXT). 
One of the advantages of high frame-rate imaging is that it is possible to achieve a high time-resolution by combining the single exposures into short-exposure images. This technique could be applied to V11 to determine the
time scale of the optical flickering, but it is too faint to be able to do this with our data. 

\begin{table*}
\caption{NGC~6652: Details of the 1 known (V11) and 2 new (V13, V14) variables in our FoV.}
\label{table:NGC6652_V}      
\centering                          
\begin{tabular}{c c c c c c c c c c c}        
\hline\hline                 
Var & RA (J2000.0) & Dec. (J2000.0) & Epoch $(d)$ & P $(d)$ & $< I >$ & $A_{i'+z'}$ & Blend & Classification\\    
\hline                        
V11 & 18:35:44.551 & -32:59:38.38 &  -  &  -  & 18.72 & 1.09 & iii & LMXB?, VFXT?\tablefootmark{a}  \\
V13 & 18:35:45.805 & -32:59:35.94 &  -  &  -  & 14.17 & 0.04 & iii & ?  \\
V14 & 18:35:46.325 & -32:59:32.79 & 6506.6618 & 0.189845 & 17.74 & 0.29 & iii & RR1?, EW?  \\
\hline                                   
\end{tabular}
\tablefoot{\varTableText{2456458.84}{}
\tablefoottext{a}{Classification from \citet{Stacey2012}}}
\end{table*}

\subsubsection{New variables}
We found two new variables in the cluster:
\begin{description}
\item[{\bf V13:}]{This star has the same position as the X-ray source H, and we therefore assume that it is the same star. 
The star has a light curve scatter of less than $\sim$0.01 mag consistent with noise. However, at the epochs HJD 2456506 and 2456509 it is $\sim$ 2 and 4\% brighter than the base-line, suggesting that the star has undergone an outburst. 
However, the CMD puts the star on the RGB, where instead we might expect long period semi-regular variability, but this seems not to be the case for V13.
We therefore refrain from classifying this variable.}
\item[{\bf V14:}]{Based on its position in the CMD this star could be a blue-straggler eclipsing binary. 
The phased light curve looks like that of a RR1 star, but the period is too short for this. Furthermore, with its position $\sim$2.5 magnitudes below the horizontal branch, it would be a field RR1 star lying behind the cluster.}
\end{description}


\section{Discussion} \label{sec:discussion}

\subsection{NGC~6388 and NGC~6441} \label{sec:disc6388_6441}
Of the five clusters that we studied in this paper, NGC~6388 and NGC~6441 contain by far the largest number of variable stars. 
These two clusters have a very similar metallicity and central concentration (see Table~\ref{table:NGCs}) and very similar CMDs (see Fig.~\ref{fig:NGC6388_cmd}, \ref{fig:NGC6441_cmd}).
However, before the present study, NGC~6441 had twice as many reported variables as NGC~6388. 

We were able to find  $\sim50$ new variables in each cluster, of which at least three are RRL stars and at least five are CW stars. 
The fact that we were able to find new short period variable stars in NGC~6441 is somewhat surprising given that a HST snapshot study has been performed on this cluster. 
It should be noted that the HST data were analysed using the DAOPHOT/ALLSTAR/ALLFRAME routines \citep[see][]{Stetson1987,Stetson1994}, which might not perform as well as DIA in such a crowded field.

The RR0 star V151 in NGC~6441 has a very large amplitude and the reason why this star has not been detected before is most likely due to its proximity to three other variable stars, which are all several magnitudes brighter. 
That we are able to detect stars like this
implies that there might be several other RRL stars situated close to bright stars outside our FoV that are not detected yet. 
Our method does, however, heighten the chances that we now have a complete census of the RRL stars in the central regions of NGC~6388 and NGC~6441.

With the newly found RRL stars there are now 23 RRL stars in NGC~6388 with robust classification (excluding V74 from our work), of which 11 are RR0 and 12 are RR1.
Using these RRLs we calculate mean periods for the RR0 and RR1 stars as $<P_{RR0}> = 0.700$ d and $<P_{RR1}> = 0.389$ d, respectively, and the ratio of RR1s to the total number of RRLs as $\frac{n_{RR1}}{n_{RRL}} \approx 0.52$.

For NGC~6441 the total number of RRL stars with robust classification is now 69 (excluding V152 from our work), of which 45 are RR0, 23 are RR1, and 1 is RR01.
From these we calculate $<P_{RR0}> = 0.745$ d, $<P_{RR1}> = 0.368$ d, and $\frac{n_{RR1}}{n_{RRL}} \approx 0.34$.

\citet{Oosterhoff1939} called attention to a relation between the mean periods and relative proportions of RR0 and RR1 stars in GCs. Generally it is found that the RRLs in Oosterhoff type I (Oo I) clusters have shorter mean periods, and a lower ratio of RR1 stars, than in Oosterhoff type II (Oo II) clusters. 
It has also been found that Oo I clusters are usually more metal rich than Oo II clusters \citep{Smith1995}. 
NGC~6388 and NGC~6441 are therefore unusual as they do not fit into either of the Oo types. 
Their high metallicity indicates that they are Oo I, but the long mean periods of the RRLs and the high ratio of RR1 stars, especially for NGC~6388, indicates Oo II. 
Based on this \citet{Pritzl2000} suggested creating a new Osterhoff type III, but \citet{Clement2001} found that the period-amplitude relation of the RR0 variables in NGC~6441 is more consistent with that of Oo II.
Our new variable star discoveries have not changed the Oo classification situation for these clusters.

\begin{figure}[t]
   \centering
   \includegraphics[width=\linewidth]{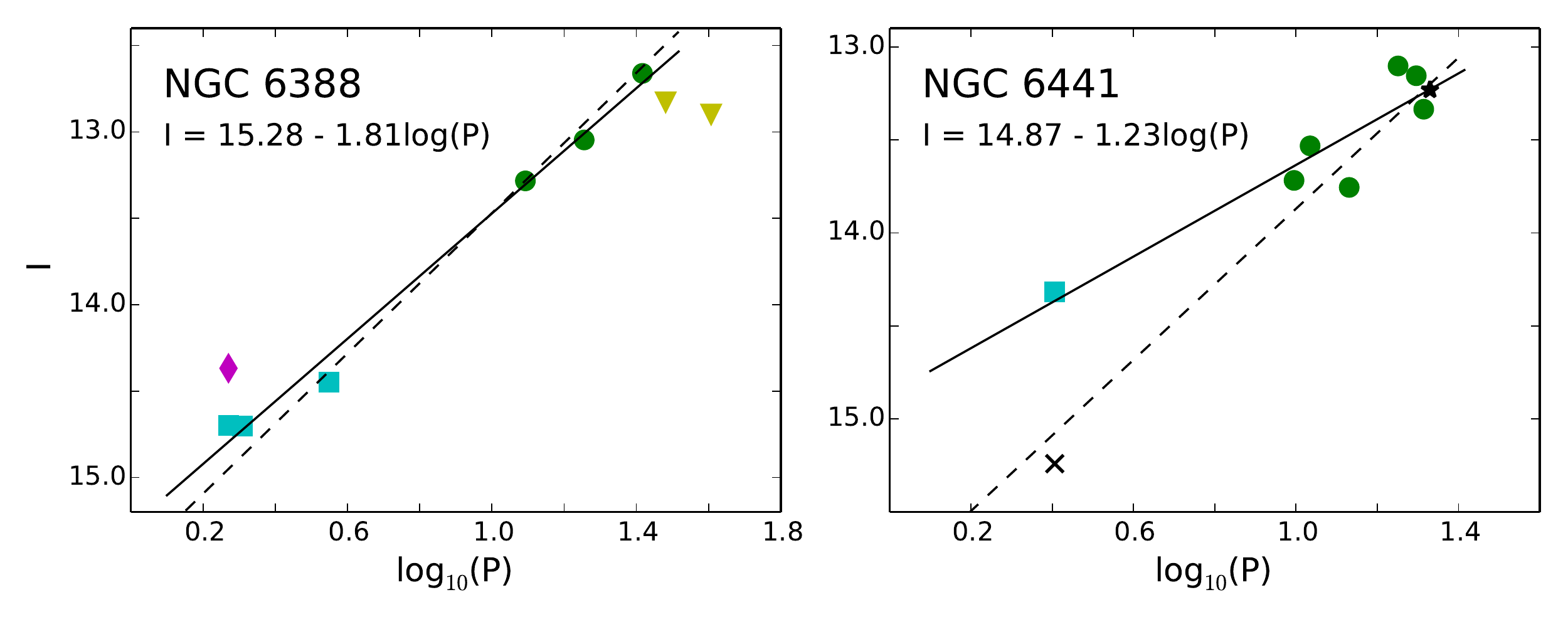}
      \caption{Plot of periods against $I$ magnitudes for the CWs and RVs found in NGC~6388 and NGC~6441. Green circles are CWAs, cyan squares are CWBs, yellow triangles are RVs, and magenta diamonds are ACs. The black star and black cross in the NGC~6441 plot are the variable stars V6 and V132, respectively, for which mean $I$ magnitudes and periods are taken from P03. 
      The solid lines are derived based on CWAs and CWBs from this study only and have $R^2$-values of 0.991 and 0.848 for NGC~6388 and NGC~6441, respectively. The dashed lines are the relations found in P03 (see Sect. \ref{sec:disc6388_6441} for details).}
        \label{fig:PerLum}
   \end{figure}

Period-luminosity diagrams for the CW and RV stars that are found in the two clusters are shown in Fig. \ref{fig:PerLum}. 
In P03 the relation between the absolute $I$ magnitudes ($I_{\text{abs}}$) of the CW stars and their periods is found to be $I_{\text{abs}} \propto - 2.03\cdot \mathrm{log}_{10}(P)$. 
The relation is actually only based on CW stars from NGC~6441, but we have plotted it for both clusters in Fig. \ref{fig:PerLum} (dashed line) using the relation $I = I_0 - 2.03 \cdot \mathrm{log}_{10}(P)$, where $I_0$ is 15.5 and 15.9 mag for NGC~6388 and NGC~6441, respectively.

For NGC~6388, there are three CWs, (V29, V63, and V77) with periods below 3 days. V29 and V63 have been classified as CWB, while V77 has been classified as an AC due to the fact that it is a little too bright for its period compared with the other two \citep{Clement2001}.
The two stars with periods over 30 days are the RV stars V80, and V82. RV stars also exhibit a correlation between their luminosity and periods, but with a larger intrinsic scatter than the CW stars, which is also evident for NGC~6388.
We find a relation of $I \propto 1.81 \cdot \mathrm{log}_{10}(P)$ for NGC~6388, which is close to the relation found in P03. 

In the P03 analysis of NGC~6441 there is one CWA star that is not included in this study. The star, V6, is classified as P2C in P03 and is included in the NGC~6441 plot in Fig. \ref{fig:PerLum}.
The CWB star V132 is heavily blended and this is probably the reason why the pipeline has measured the star to be a magnitude brighter than in P03. 
Using our $I$ magnitude for V132, we find the relation $I \propto 1.23 \cdot \mathrm{log}_{10}(P)$, which is quite far from the P03 relation. However, by using the $I$ magnitude from P03 for V132 (plotted as a black cross in Fig. \ref{fig:PerLum}), the relation found in P03 looks much more feasible.

Most of the new variables found in the two clusters are RGB stars, which is not surprising due to a number of factors: 
\begin{itemize}
\item{The CMDs of the two clusters show a very prominent RGB, on which most stars are intrinsically variable \citep[e.g.][]{Percy2007,Percy2014}.}
\item{The baseline of our data is about 14 months, which makes it possible to detect variability over a long period.}
\item{The filter we use goes from the optical red to near-infrared wavelengths, which matches very well with the colour of the RGB stars.}
\item{The use of high frame-rate imaging makes it possible to observe much brighter stars without saturating. }
\end{itemize}

Both the SR and L classifications have several subclasses, but in order to safely assign these, one would need more information on the stars, such as spectral type and a longer baseline than is available in this study.

\subsection{NGC~6638}
Of the 21 variable stars in R77 where a period has been determined, one (V24) has a period consistent with that of an RR0, and 14 with RR1 periods. 
Two of the RR1s (V40 and V42) are located very far from the centre and might not be cluster members. 
We find that a further eight of the known variables (those in Table \ref{table:NGC6388_V}) have periods consistent with RRLs and we also retrieve five new RRL stars (Table \ref{table:NGC6388_NV}). 

\begin{figure}[t]
   \centering
   \includegraphics[width=\linewidth]{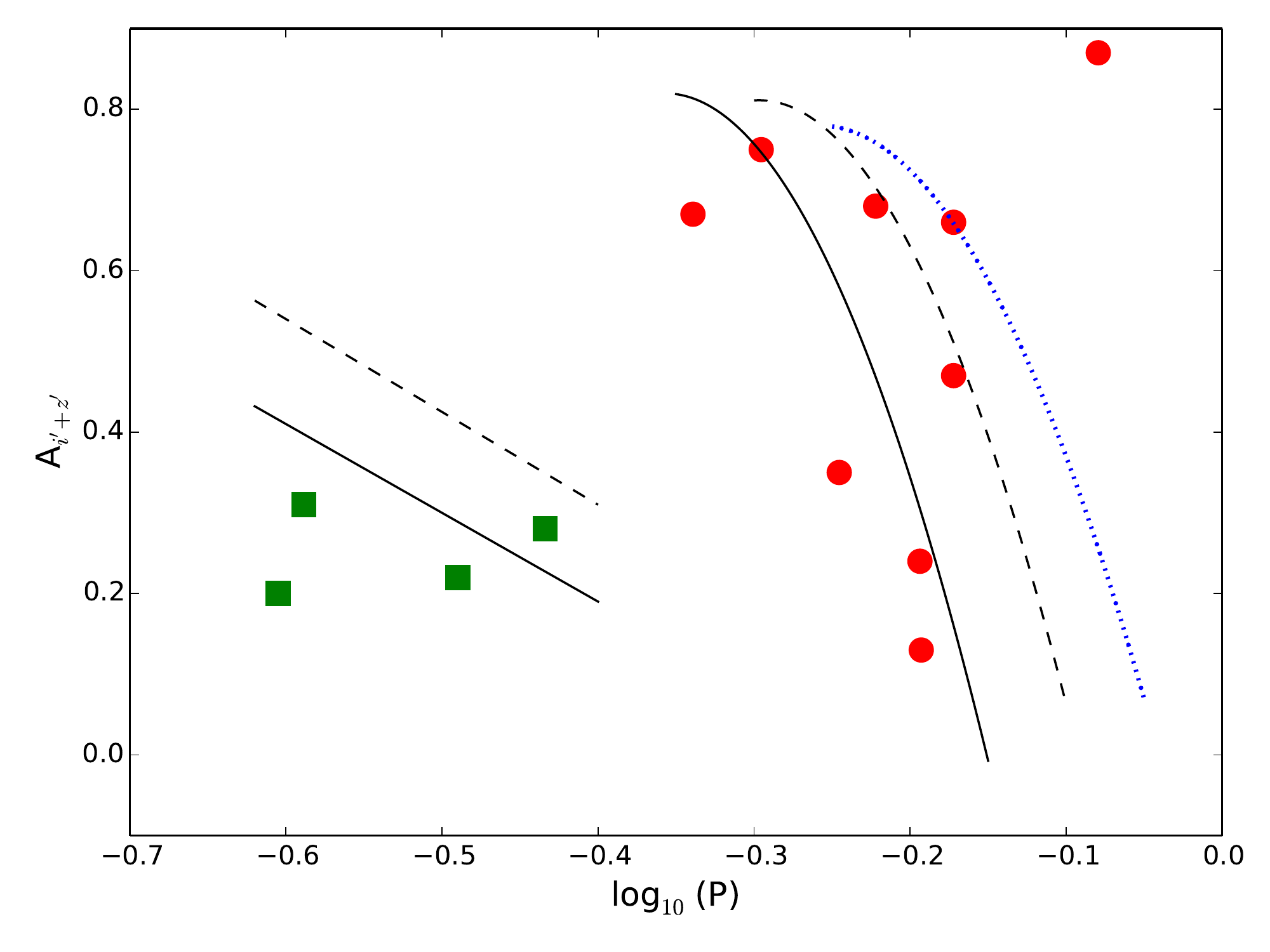}
      \caption{Bailey diagram of NGC~6638 for $I$ amplitudes, with RR0 and RR1 stars plotted as red circles and green squares, respectively. The black solid and dashed lines are the Oo I and Oo II loci, respectively, as calculated by \citet{Kunder2013} for $I$-band. The blue dotted line is the locus found by \citet{ArellanoFerro2011, ArellanoFerro2013a} for the two Oo II clusters NGC 5024 and NGC 6333, respectively. }
        \label{fig:Bailey}
   \end{figure}

The total number of RRL stars in the cluster is therefore between 26 and 28, of which 10 are RR0 and 16-18 are RR1. 
The mean periods have been calculated as $<P_{RR0}> = 0.625$ d and $<P_{RR1}> = 0.305$ d excluding V40 and V42, and including them only changes this by less than 0.001 d. 
The ratio of RR1s to total number of RRLs is between 0.62 and 0.64, depending on the inclusion or exclusion of V40 and V42.

Fig. \ref{fig:Bailey} shows a Bailey diagram, i.e. a plot of the period versus the amplitude for RR Lyrae stars, for NGC~6638. The black lines in this plot are the loci calculated by \citet{Kunder2013} for the Oo I clusters for the RR0 and RR1 stars (solid lines) and for the Oo II clusters (dashed lines).
The dotted blue line is the locus found by \citet{ArellanoFerro2011, ArellanoFerro2013a} for the two Oo II clusters NGC 5024 and NGC 6333, respectively.
Both sets of loci are derived from the $I$-band light curves and as our $i'+z'$ filter is slightly redder, our amplitudes might also be a little smaller than true $I$-band amplitudes.
Even with slightly larger amplitudes than what is plotted in Fig. \ref{fig:Bailey}, the RRL stars seem to follow the loci for Oo I clusters. 
There are a number of RR0 stars that lie somewhat higher than the rest, and these are probably more evolved stars, following the argumentation in \citet{Clement1997,Cacciari2005}.
Based on the Bailey diagram and the metallicity of the cluster, [Fe/H] $\simeq -0.95$, we tentatively classify NGC~6638 as an Oo I cluster, despite the high $\frac{n_{RR1}}{n_{RRL}}$ ratio. 

Two of the newly found RRLs (V64 and V68) are found very close to the same bright star, and this is a very clear example of the advantages of the EMCCD data, as these would have been very hard to resolve using conventional imaging. 
It also makes it more likely that we now have a complete census of the RRLs in the central parts of this cluster, especially as this cluster does not have a very dense central region (see Table~\ref{table:NGCs}). 

A CMD of NGC~6638 has been published by \citet{Piotto2002}, and from this it is evident that the RGB of this cluster is much less prominent than the ones of NGC~6388 and NGC~6441.
It is therefore not surprising that only very few variable RGB stars are found in this cluster.

\subsection{NGC~6528 and NGC~6652}
NGC~6528 and NGC~6652 have significantly different metallicities, but they share their lack of variable stars. 

Until now no variable stars had been reported for NGC~6528, but we were able to find two RRLs and four long period RGB stars, plus a very faint variable star that could not be classified. 

Of the three variable stars we find in NGC~6652, two seem to be optical counterparts of X-ray sources, and
the third might be a blue-straggler eclipsing binary, or a field RR1 lying behind the cluster. 

The CMDs of the two clusters (a CMD of NGC~6528 can be found in \citet{Feltzing2002}) correlate well with the number of variables found. 
Both clusters have very few stars in the instability strip of the horizontal branch. NGC~6652 has very few stars on the RGB, while the RGB of NGC~6528 is comparable to that of NGC~6388.


\section{Conclusions} \label{sec:conclusion}
A detailed variability study of 5 metal rich ([Fe/H] $>$ -1) globular clusters; NGC~6388, NGC~6441, NGC~6528, NGC~6638, and NGC~6652; has been performed, by using EMCCD observations with DIA.
All previously known variable stars located within our field of view in each of the 5 clusters have been recovered and classified, and numerous previously unknown variables have been discovered. 



For three of the clusters; NGC~6388, NGC~6441, and NGC~6652; electronically available CMD data from a HST survey have helped in the classification of the variable stars. 
For the two remaining clusters CMDs exist in the literature.
Common for all of the clusters is that the CMDs seem to be in agreement with the number and types of variable stars we have found. 

With this article we have further demonstrated the power of using EMCCDs for high-precision time-series photometry in crowded fields and it demonstrates the feasibility of large-scale observing campaigns using these methods.

\begin{acknowledgements}
The authors would like to thank Christine Clement for giving invaluable advice on variable star numbering and nomenclature, and Horace Smith for input on RR01 stars.

The operation of the Danish 1.54m telescope is financed by a grant to UGJ from the Danish Natural Science Research Council. 
We also acknowledge support from the Center of Excellence Centre for Star and Planet Formation (StarPlan) funded by The Danish National Research Foundation.
DMB and KAA acknowledge support from NPRP grant \# X-019-1-006 from the Qatar National Research Fund (a member of Qatar Foundation).
AAF acknowledges the support from DGAPA-UNAM grant through
project IN104612.
KAA, CL, MD, KH and MH acknowledge grant NPRP-09-476-1-78 from the Qatar National Research Fund (a member of Qatar Foundation).
CS has received funding from the European Union Seventh Framework Programme (FP7/2007-2013) under grant agreement no. 268421.
SHG and XBW would like to thank the financial support from National Natural Science Foundation of China through grants Nos. 10373023, 10773027 and 11333006.
TH is supported by a Sapere Aude Starting Grant from The Danish Council for Independent Research.
TCH acknowledges support from the Korea Research Council of Fundamental Science \& Technology (KRCF) via the KRCF Young Scientist Research Fellowship Programme and for financial support from KASI travel grant number 2013-9-400-00.
HK acknowledges the support from the European Commission under the Marie Curie IEF Programme in FP7.
M.R. acknowledges support from FONDECYT postdoctoral fellowship Nr. 3120097.
OW (FNRS research fellow) and J Surdej acknowledge support from the Communaut\'e francaise de Belgique - Actions de recherche concert\'ees - Acad\'emie Wallonie-Europe.

Some of the data presented in this paper were obtained from the Mikulski Archive for Space Telescopes (MAST). STScI is operated by the Association of Universities for Research in Astronomy, Inc., under NASA contract NAS5-26555. Support for MAST for non-HST data is provided by the NASA Office of Space Science via grant NNX13AC07G and by other grants and contracts.
\end{acknowledgements}

\bibliographystyle{aa}
\bibliography{EMCCD_metalrich}   
  
\clearpage

\end{document}